\documentclass[aps,pra,preprint,a4paper,doublecolumn]{revtex4-1}
\usepackage{graphicx}
\usepackage{morefloats}
\usepackage{longtable}
\usepackage{color}
\usepackage{float}
\usepackage{tablefootnote}
\def\etal{{\it et al~}}
\begin{document}


\title{Partial and Total Dielectronic Recombination 
rate coefficients for W$^{73+}$ to W$^{56+}$}


\author{S. P. Preval}
\email{simon.preval@strath.ac.uk}
\author{N. R. Badnell}
\author{M. G. O'Mullane}
\affiliation{Department of Physics, University of Strathclyde, 
Glasgow G4 0NG, United Kingdom}



\date{\today}

\begin{abstract}
Dielectronic recombination (DR) is a key atomic process which affects the spectroscopic
diagnostic modelling of tungsten, most of whose ionization stages will be found somewhere
in the ITER fusion reactor: in the edge, divertor, or core plasma.
Accurate DR data is sparse while complete DR coverage is unsophisticated (e.g. average-atom
or Burgess General Formula) as illustrated by the large uncertainties which currently
exist in the tungsten ionization balance.
To this end, we present a series of partial final-state-resolved and total DR
rate coefficients for W$^{73+}$ to W$^{56+}$ Tungsten ions. This is part of a wider
effort  within {\it The Tungsten Project} to calculate accurate dielectronic recombination rate 
coefficients for the tungsten isonuclear sequence  for use in collisional-radiative 
modelling of finite-density tokamak plasmas. The recombination rate coefficients have
been calculated with {\sc autostructure} using kappa-averaged relativistic wavefunctions 
in level resolution (intermediate coupling) and configuration 
resolution (configuration average). The results are available from OPEN-ADAS according to
the {\it adf09} and {\it adf48} standard formats. Comparison with previous calculations of
total DR rate coefficients for W$^{63+}$ and W$^{56+}$ yield agreement 
to within 20\% and 10\%, respectively, at peak temperature. It is also seen that the J\"{u}ttner 
correction to the Maxwell distribution has a significant effect on the ionization balance of 
tungsten at the highest charge states, changing both the peak abundance temperatures and 
the ionization fractions of several ions.
\end{abstract}


\maketitle

\section{Introduction}
ITER \footnote{{h}ttp://www.iter.org} is posited to be the penultimate step in realizing a 
nuclear fusion power plant. It will be significantly larger than present machines, such as JET, 
in terms of plasma volume, core temperature, and physical size \cite{iter99}. 
Beryllium coated tiles will line the wall of the main reactor vessel due to their low erosion rate 
and the low tritium retention of Be. Tungsten ($Z=74$) will be used in regions of high power-loads,
such as the divertor chamber at the base of the main vessel, and it is also 
resistant to tritiation \cite{kharbachi14a}. On the downside, such high-$Z$ elements
are efficient radiators and must be kept to a minimum in the main plasma to avoid
degrading its confinement. Because of this, JET has undergone a major upgrade to an 
ITER-like wall to act as a test-bed. Control of tungsten sources and its subsequent 
transport are under intensive study \cite{fedorczak15a}.
Tungsten is the highest-$Z$ metal in widespread use in a tokamak. 
Prior to the installation of the ITER-like wall at JET, 
molybdenum ($Z=42$) was the highest-$Z$ metal in widespread use, at Alcator C-Mod. 
Like tungsten, molybdenum has a low tritium absorption rate \cite{zakharov1975a}. However, 
molybdenum has a significantly lower melting point than tungsten, and also transmutes to technetium, 
complicating reactor decommissioning.

\begin{figure}
\begin{centering}
\includegraphics[width=120mm]{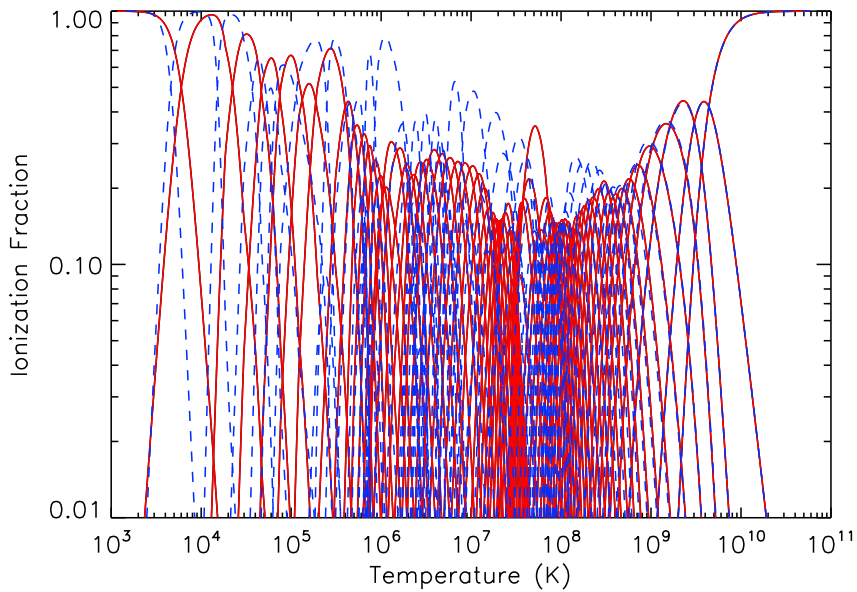}
\caption{Zero-density fractional abundances of tungsten ionization stages,
calculated using P\"{u}tterich~\etal \cite{putterich08a} recombination data 
(red, solid curves) and Foster \cite{foster08a} recombination data
(blue, dashed curves). Both use the ionization rate coefficients 
from Loch \etal \protect\cite{loch05a}.}
\label{fig:fostputtcomp}
\end{centering}
\end{figure}

Most of the tungsten isonuclear sequence needs to be covered by non-LTE plasma modelling,
from its initial sputtering from surfaces through the edge, divertor and core plasmas.
One of the most basic quatities is the tungsten ionization balance: a measure of
the dominant ionization stages as a function of temperature and density. While our
understanding of the ionization rates required appears to be in reasonable order 
\cite{loch05a}, the same is not true for the competing dielectronic plus radiative
recombination rates (DR+RR). In Figure~\ref{fig:fostputtcomp} we compare the zero-density
ionization balance for tungsten obtained using two different sets of recombination data
\cite{putterich08a,foster08a} and the same ionization rate coefficients, 
of \cite{loch05a}. It can be seen that there are large discrepancies between the peak temperatures 
of individual ionization stages, as well as the fractional population of said ionization stage. 
The dielectronic recombination data of P\"{u}tterich~\etal \cite{putterich08a} was calculated with 
ADPAK \cite{post1977a,post1995a} using an average-atom model and it was scaled by the authors 
in order to improve agreement between theory and experiment with regards to the shape of the fractional 
abundances of W$^{22+}$--W$^{55+}$. The DR data of Foster \cite{foster08a}
used the Burgess General Formula \cite{burgess1965a}. Both used the same scaled hydrogenic radiative
recombination data. Clearly, more reliable DR data is required.

Another issue is that the magnetic fusion plasmas cannot be taken to be a zero-density
one. The true ionization balance is density dependent and the corresponding
density dependent (effective) ionization and recombination rate coefficients
are obtained from collisional-radiative modelling.
The ionization rate coefficients are much less sensitive to density effects than the 
recombination ones because dielectronic recombination takes place through and
to high-Rydberg states. Therefore, partial final-state-resolved rate coefficients are needed.
Even where detailed calculations have been made, the data available is usually
in the form of zero-density totals, i.e. summed-over all final states. As such,
it is difficult to use such data for collisional-radiative modelling in reliable manner.

Detailed calculations have been performed for a select few ions of tungsten. However, these 
are very sparse, and tend to be for closed-shell ions which are important for plasma 
diagnostics. The most complicated exception to this to date is our work 
on the open f-shell: W$^{20+,18+}$ (4$d^{10}$ 4$f^{q}, q=8,10$)\cite{badnell12a,spruck14a}.  
Data for these ions were calculated using an upgraded version of {\sc autostructure} designed to 
handle the increased complexity of the problem. The {\sc hullac} \cite{barshalom01a} and the Cowan code 
\cite{cowanbook81} have been used by Safronova~\etal
\cite{usafronova12a,usafronova11a,usafronova12b,usafronova09a,usafronova09b} to calculate
 DR rate coefficients for W$^{5+}$, W$^{28+}$, W$^{46+}$, W$^{63+}$, and W$^{64+}$,
respectively. Behar~\etal \cite{behar99a} and Peleg~\etal \cite{peleg98a} have also
used these codes for W$^{46+}$,W$^{64+}$, and W$^{56+}$, respectively. In addition, the 
Flexible Atomic Code ({\sc fac}, \cite{gu03a}) has been used by Meng~\etal \cite{meng09a}
and Li~\etal \cite{li12a} to calculate DR rate coefficients for W$^{47+}$ and W$^{29+}$, respectively.
Just recently, Wu~\etal \cite{wu15a} have calculated zero-density total DR rate coefficients 
for W$^{37+}$ -- W$^{46+}$ using {\sc fac}.

In contrast, partial RR rate coefficients have been calculated for the entire isonuclear sequence 
of Tungsten, and the results presented in a series of papers, by Trzhaskovskaya~\etal
\cite{trzh10a,trzh13a,trzh14a,trzh14b}. The authors used a Dirac-Fock method with fully 
relativistic wavefunctions, and included contributions from all significant radiation multipoles. 
The authors state that the majority of their RR rate coefficients were calculated to $<1$\% numerical accuracy. 
However, for outer shell RR and high temperatures, their rate coefficients were calculated to 
$<5$\% \cite{trzh10a}. The authors also present total RR rate coefficients summed up to $n=20$ and $\ell=19$.

In order to address this situation, we have embarked on a programme of work, as part of 
{\it The Tungsten Project}, which aims to calculate partial final-state resolved DR rate 
coefficients for use in  collisional-radiative modelling with ADAS \footnote{{h}ttp://www.adas.ac.uk}
for the entire isonuclear sequence of tungsten.
For completeness and ease of integration within ADAS, we compute the corresponding RR
data at the same time. Zero-density totals are readily obtained from the archived data.
The work presented here covers W$^{73+}$ to W$^{56+}$.

On a practical technical point, the names of various elements in the periodic table 
are not particularly helpful to label ionization stages of a
large isonuclear sequence such as tungsten. Thus, we will not refer to such species by a
name such as Pr-like. Instead, we adopt a notation based on the number of electrons 
possessed by a particular ion. For example, H-like (1 electron) W$^{73+}$ will be referred to as 
01-like, Ne-like (10 electrons) W$^{64+}$ will be referred to as 10-like, and Pr-like (59 electrons)
W$^{15+}$ as 59-like. This mirrors our database archiving.

The outline of this paper is as follows: in Sec. II we outline the background theory for our
description of DR and RR, as encapsulated in the {\sc autostructure} code, and give consideration
to the delivery of data in a manner appropriate for collisional-radiative modelling.
In Sec. III, we describe our calculations for 00-like to 18-like ions. In Sec. IV, we present our results
for DR/RR rate coefficients and compare them with those published previously, where available; 
then we look at how the zero-density ionization balance of tungsten changes on using our new recombination data.
We conclude with some final remarks, and outline future calculations.

\section{Theory}
We use the distorted-wave atomic package {\sc autostructure} \cite{badnell86a,badnell97a,badnell11a}.
For recombination, {\sc autostructure} makes use of the independent processes
and isolated resonance approximations \cite{pindzola92a}. Then, the partial DR rate coefficient 
$^{DR}\alpha_{f\nu}^{z+1}$, from some initial state $\nu$ of ion $X^{+z+1}$ to a final state 
$f$ of ion $X^{+z}$, can be written as
\begin{equation}
^{DR}\alpha_{f\nu}^{z+1}(T_e)= \left(\frac{4{\pi}a_{0}^{2}I_{H}}{k_{B}T_{e}}\right)^{\frac{3}{2}}
\sum_{j}\frac{\omega_{j}}{2\omega_{\nu}}\exp{\left[-\frac{E}{k_{B}T_{e}}\right]}
\frac{\sum_{l}{A_{j\rightarrow{\nu},E\,l}^a}{A_{j\rightarrow{f}}^{r}}}
{\sum_{h}{A_{j\rightarrow{h}}^{r}} + \sum_{m,l}{A_{j\rightarrow{m},E\,l}^{a}}},
\label{DReq}
\end{equation}
where the $A^{a}$ are the autoionization rates, $A^{r}$ are the radiative rates, $\omega_{j}$ 
is the statistical weight of the $N$-electron target ion, $E$ is the total energy of the 
continuum electron, minus its rest energy, and with corresponding orbital angular momentum 
quantum number $l$ labelling said channels. $I_{H}$ is the ionization energy of the 
hydrogen atom, $k_{B}$ is the Boltzmann constant, $T_{e}$ is the electron temperature, and 
$(4{\pi}a_{0}^{2})^{3/2}=6.6011\times{10}^{-24}$cm$^{3}$. The sum over the autoionizing states $j$ 
recognizes the fact that, in general, these states have sufficiently short lifetimes in
a magnetic fusion plasma for them not be collisionally redistributed
before breaking-up, although statistical redistribution is assumed in some cases \cite{badnell06a}.

The partial RR rate coefficient  $^{RR}\alpha_{f\nu}^{z+1}$ can be written, in terms of the 
photoionization cross section $^{PI}\sigma_{\nu f}^{z}$ for the inverse process using detailed balance, as
\begin{equation}
^{RR}\alpha_{f\nu}^{z+1}(T_e) = \frac{c\,\alpha^3}{\sqrt{\pi}}
\frac{\omega_{f}}{2\omega_{\nu}}\left(I_H k_B T_e\right)^{-3/2}
\int^\infty_0 E^2_{\nu f}\, {^{PI}\sigma_{\nu f}^{z}(E)}
\exp{\left[-\frac{E}{k_{B}T_{e}}\right]}dE\,,
\label{RReq}
\end{equation}
where $E_{\nu f}$ is the corresponding photon energy and $c\,\alpha^3/\sqrt{\pi}=6572.67$cm\,s$^{-1}$
for $^{PI}\sigma_{\nu f}^{z}$ given in cm$^{2}$. 
The photoionization cross sections for arbitrary electric and magnetic multipoles are 
given by \cite{grant74a}. The numerical approaches to converging the quadrature accurately and 
efficiently have been given previously \cite{badnell06a}.

At high temperatures ($\gtrsim 10^9$~K) relativistic corrections to the usual
Maxwell-Boltzmann distribution become important. The resultant Maxwell-J\"{u}ttner distribution 
\cite{synge57a} reduces simply to an extra multiplicative factor, $F_{\mathrm{r}}(\theta)$, 
to be applied to the Maxwell-Boltzmann partial rate coefficients:
\begin{equation}
F_{\mathrm{r}}(\theta) = \sqrt{\frac{\pi\theta}{2}} 
\frac{1}{K_{2}(1/\theta){\rm e}^{1/\theta}},
\end{equation}
where $\theta=\alpha^2 k_{B}T/2I_H$, $\alpha$ is the fine-structure constant and $K_{2}$
is the modified Bessel function of the second kind.
This factor is normally consistently omitted from data archived in ADAS, being subsequently 
applied if required in extreme cases. However, since it has a non-negligible affect at
the temperature of peak abundance for the highest charge states we consistently
include it {\it for all} tungsten DR and RR data and flag this in the archived files. 

Plasma densities in magnetic fusion reactors vary greatly. For ITER, the plasma densities are thought to vary 
from $10^{10}$--$10^{13}$~cm$^{-3}$ for the edge plasma, to $\sim{10}^{14}$~cm$^{-3}$ for the core plasma, 
reaching $\sim{10}^{15}$~cm$^{-3}$ for the divertor plasma. Because of these densities, the coronal 
picture breaks down: capture into an excited state does not cascade
down to the ground uninterrupted. Instead, further collisions take place, leading
in particular to stepwise ionization, for example. This strongly suppresses
coronal total recombination rate coefficients. Collisional-radiative (CR) modelling
of the excited-state population distribution is necessary. This leads to
density-dependent effective ionization and recombination rate coefficients.
A key ingredient for CR modelling is partial {\it final-state-resolved}
recombination data. Our approach for light systems is detailed in 
\cite{badnell03a} and \cite{badnell06a} for DR and RR, respectively.
Low-lying final-states are fully level-resolved while higher-lying states
are progressively ($n\ell$- and $n$-) bundled over their total quantum numbers,
whilst retaining their level parentage. Initial ground and metastable levels
are also fully-resolved. The data are archived in ADAS standard formats, 
viz. {\it adf09} (DR) and {\it adf48} (RR). One does not need to progress far into the M-shell
for the number of such final states to become unmanageable by CR modelling and
further bundling is required. This is carried-out most efficiently as the partial
recombination rate coefficients are calculated and leads to much more compact {\it adf09} and {\it adf48} files.
We find it necessary to bundle over all final recombined levels within a configuration.
For such configurations which straddle the ionization limit we include the
statistical fractions within the {\it adf} files. The initial ground and metastable
levels remain level-resolved, as does the calculation of autoionizing branching
ratios (fluorescence yields). We describe such a mixed resolution scheme as
a `hybrid' approach and the {\it adf} files are labelled accordingly.
All resultant {\it adf09} and {\it adf48} files are made available via 
OPEN-ADAS \footnote{{h}ttp://open.adas.ac.uk/}.

\section{Calculations}
All rates and cross sections were determined on solving the kappa-averaged quasi-one-electron
Dirac equation for the large and small components utilizing the Thomas-Fermi-Dirac-Amaldi 
model potential \cite{eissner69a} with unit scaling
parameters to represent the $N$- and $(N+1)$-electron ions.
We utilized several coupling schemes. Configuration average (CA) was used to give a quick
overview of the problem. This neglects configuration mixing and relativistic interactions
in the Hamiltonian. $LS$-coupling (LS) allows for configuration mixing but tends to
overestimate it in such highly-charged ions because relativistic interactions push 
interacting terms further apart. Thus, our main body of data is calculated in
intermediate coupling (IC). For the K-shell ions, we included valence-valence two-body 
fine structure interactions. These gave rise to a $\sim$7\% increase in the total DR rate 
coefficients for 01-like and 02-like ions at high temperatures. 
We neglect these interactions for the L- and M-shell ions
since the increase in the total DR rate coefficient is $<1$\%. 

\subsection{DR}
It is necessary to include all dominant DR reactions illustrated by Eq.(\ref{DReq}).
The initial state $\nu$ is taken to be the ground state. Metastables are unlikely
to be important at such high charge states. The driving reactions are the autoionizing 
states produced by one-electron promotions from the ground configuration, with a 
corresponding capture of the continuum electron. We label these core-excitations by the 
initial ($n$) and final ($n'$) principal quantum numbers of the promoted electron, 
and include all corresponding sub-shells ($\ell$-values). 
The dominant contributions come from $\Delta n=0$ 
($n=n'$) and $\Delta n=1$ ($n'=n+1$), being well separated in energy/temperature.
Contributions from $\Delta n>1$ tend to be suppressed by autoionization into excited
states, as represented by the sum over $A^a$ in the denominator of (\ref{DReq}).
The outermost shell dominates but the $\Delta n=1$ inner-shell promotion ($n=n'-1$)
can be significant when there are few outer $n$-shell electrons. As their number increases,
core re-arranging autoionizing transitions suppress this inner-shell contribution.
These core-excitations define a set of $N$-electron configurations to which 
continuum and Rydberg electrons are coupled.

Based-on these promotion rules, the core-excitations considered for each ion (W$^{73+}$ to W$^{56+}$)
are listed  in Table \ref{table:drcorex}. The calculations were carried-out first in CA to determine
which excitations are dominant. We omitted core-excitations that contribute $<1$\% to the sum 
total of all DR core-excitation rate coefficients spanning the ADAS temperature grid. This grid 
covers $10z^2$--$2\times{10}^{6}z^2$\,K, where $z$ is the residual charge of the initial target ion. 
DR for the dominant core-excitations is then calculated 
in IC. The $n\ell$ Rydberg electron, in the sum over autoionizing states $j$, is calculated explicitly 
for each principal quantum number up to $n=25$ and then on a quasi-logarithmic $n$-mesh up to $n=999$.
The partial DR rate coefficient tabulation is based on this mesh of $n$-values.
Total (zero-density) DR rate coefficients are obtained by interpolation and quadrature
of these partials.
The maximum Rydberg orbital angular momentum ($\ell$) is taken to be 
such that the total rate coefficients are converged to better than 1\% over the ADAS temperature range.
Radiative transitions of the Rydberg electron to final states with principal
quantum number greater than the core-excitations' are described hydrogenically.
Those into the core are described by a set of $(N+1)$-electron configurations
which are generated by adding another core electron orbital to all $N$-electron
configurations describing the core-excitations. In the case of $\Delta n>1$
core-excitations this also allows for dielectronic capture into the core.

To make clear to the complete set of configurations included for a typical
calculation, we give a list of configurations used to calculate DR rate 
coefficients for 14-like $3-3$ and $3-4$ core-excitations in Table \ref{table:14likeconf}.
We have marked also, with an *, configurations which were added to allow for the dominant
configuration mixing within a complex by way of the `one up, one down rule'. For example, 
the configuration $3s 3p^2 3d$ strongly mixes with $3p^4$ and $3s^2 3d^2$.

\subsection{RR}
The partial RR rate coefficients were calculated in a similar, but simplified fashion,
to $\Delta{n}=0$ DR, viz., the $N$-electron target configurations were restricted to
those which mixed with the ground and the $(N+1)$-electron configurations 
were these $N$-electron configurations with an additional core electron. The Rydberg $n\ell$-values
were again calculated for each $n$ up to $n=25$ and then on the same $n$-mesh as used for DR, 
up to $n=999$, with $\ell=0-10$, relativistically. At high-$T$ ($>10^9$)~K, many multipoles contribute to
the photoionization/recombination at correspondingly high energies \cite{pratt1973a}.
We included up to E40 in CA and E40/M39 in the IC calculations, which is sufficient to
converge the total RR rate coefficients to $<1$\% over the ADAS temperature range.
A non-relativistic (dipole) top-up was then used to include up to
$\ell=150$ to converge the low-temperature RR rate coefficients --- relativistic effects being 
negligible there. This approach is sufficient to calculate the total RR rate coefficients to better
than 1\%, numerically.

\section{Results and Discussion}
In this section we present the results of our DR and RR rate coefficient calculations
for 00-like to 18-like. In our plots we show the tungsten fractional peak abundance curves
from P\"{u}tterich~\etal \cite{putterich08a} to give an indication of the relevant temperatures
for application purposes. At these temperatures, RR is dominated by capture into
the lowest available $nl$-subshell.
We consider the DR rate coefficients first, and look at the K, L, and M shells in turn. 
Next, we consider the RR rate coefficients, and assess their importance relative to DR. 
We compare our results with others, where possible.
Finally, we look at the effect on the zero-density ionization balance of tungsten 
when using our new data.

\subsection{K-shell DR}
\begin{figure}
\begin{centering}
\includegraphics[width=120mm]{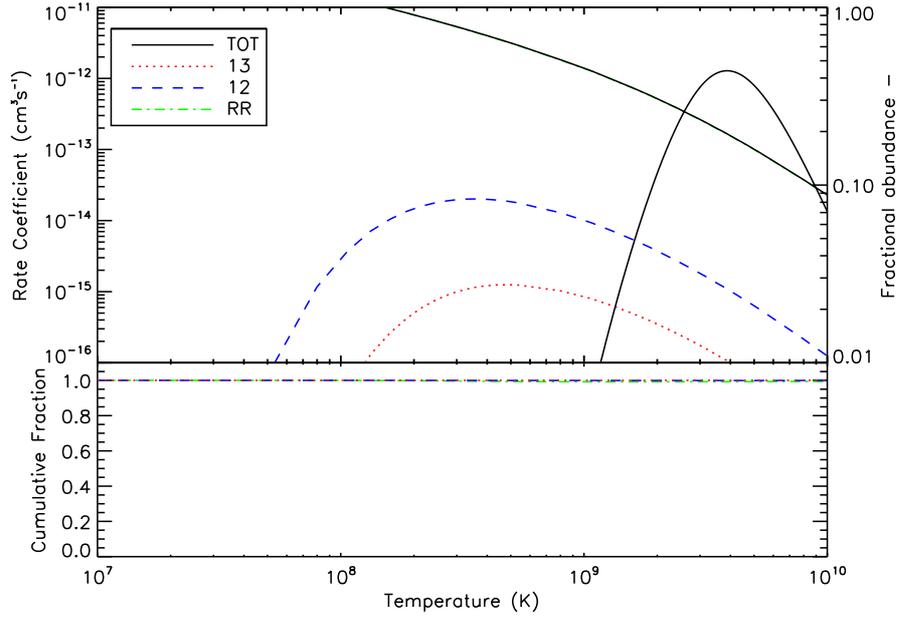}
\caption{01-like DR rate coefficients for core-excitations 1-2 and 1-3, along
with the RR rate coefficient, and the sum total of these. The solid black curve is the fractional abundance
for 01-like as calculated using P\"{u}tterich~\etal's \cite{putterich08a} scaled recombination data, 
and Loch~\etal's \cite{loch05a} ionization data. The bottom subfigure is the cumulative sum of these different contributions 
(see text for an explanation).}
\label{fig:hlikeicr}
\end{centering}
\end{figure}

\begin{figure}
\begin{centering}
\includegraphics[width=120mm]{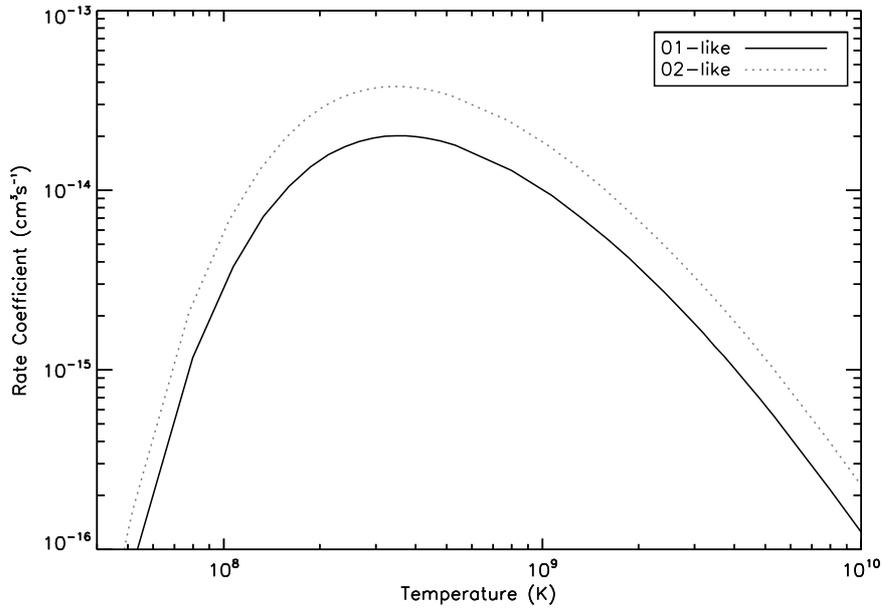}
\caption{DR rate coefficients for 1-2, $\Delta{n}=1$ core-excitation for 01- and 02-like}
\label{fig:kshelld112}
\end{centering}
\end{figure}

The DR rate coefficients for 01 and 02-like are very small compared to RR. The reason for 
this is that the RR rate coefficient scales as $z$ (residual charge) while the
DR rate coefficient here scales as $z^{-3}$, being proportional to the dielectronic capture 
rate (the fluorescence yields are close to unity due to the $z^4$ scaling of the
radiative rates and the $z^0$ ($=1$) of the autoionization.) In Figure~
\ref{fig:hlikeicr} we have plotted the DR and RR rate coefficients for 01-like. In the top 
subplot, we show the individual contributions from each DR core-excitation, and RR. The ionization 
balance for 01-like, calculated using the scaled recombination data of P\"{u}tterich~\etal 
\cite{putterich08a} and the ionization data of Loch~\etal \cite{loch05a}, 
is plotted also for reference. In the bottom subplot, we have plotted the 
cumulative sum of each contribution to the total recombination rate coefficient. This was 
calculated by taking the fraction of the largest contribution to the total recombination rate 
coefficient. The next curve is calculated by adding the first and second largest
contributions together, and taking the fraction of this to the total recombination 
rate coefficient, and so on. It can be seen that 
the total recombination rate coefficient is dominated by RR, it being at least
two orders of magnitude larger than DR at any temperature of interest.
Comparatively, the DR $\Delta{n}=1$ core-excitations for 01- and 02-like are a factor 10 
larger than their corresponding $\Delta{n}=2$ core-excitations. 
This is due to the (core) $n^{-3}$ scaling of the autoionization rate, rather
than $\Delta{n}=1$ autoionization into excited states for the $\Delta{n}=2$.
Finally, in Figure~\ref{fig:kshelld112} we compare the total 01- and 02-like
DR rate coefficients. The 02-like is roughly a factor of two larger because there
are two K-shell electrons available to promote.

\subsection{L-shell DR}
\begin{figure}
\begin{centering}
\includegraphics[width=120mm]{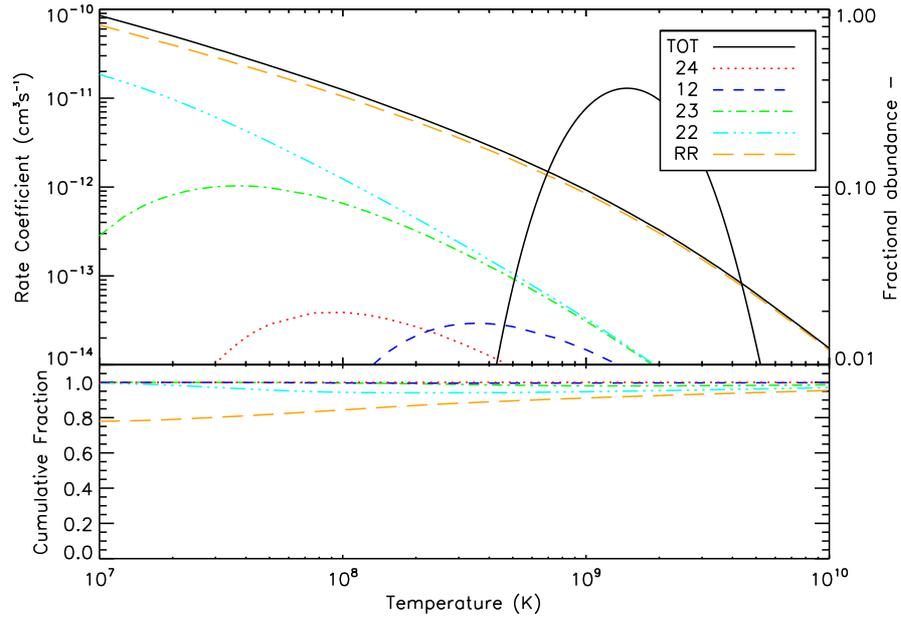}
\caption{As Figure~\ref{fig:hlikeicr}, but for 03-like with core-excitations 1-2, 2-2, 2-3, and 2-4.}
\label{fig:lilikeicr}
\end{centering}
\end{figure}

\begin{figure}
\begin{centering}
\includegraphics[width=120mm]{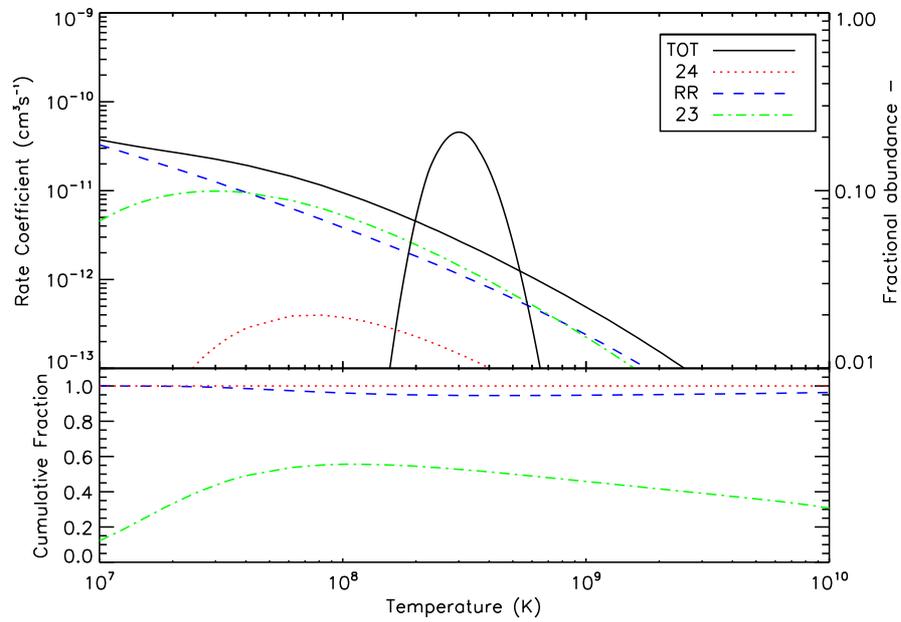}
\caption{As Figure~\ref{fig:hlikeicr}, but for 10-like with core-excitations 2-3 and 2-4.}
\label{fig:nelikeicr}
\end{centering}
\end{figure}

\begin{figure}
\begin{centering}
\includegraphics[width=120mm]{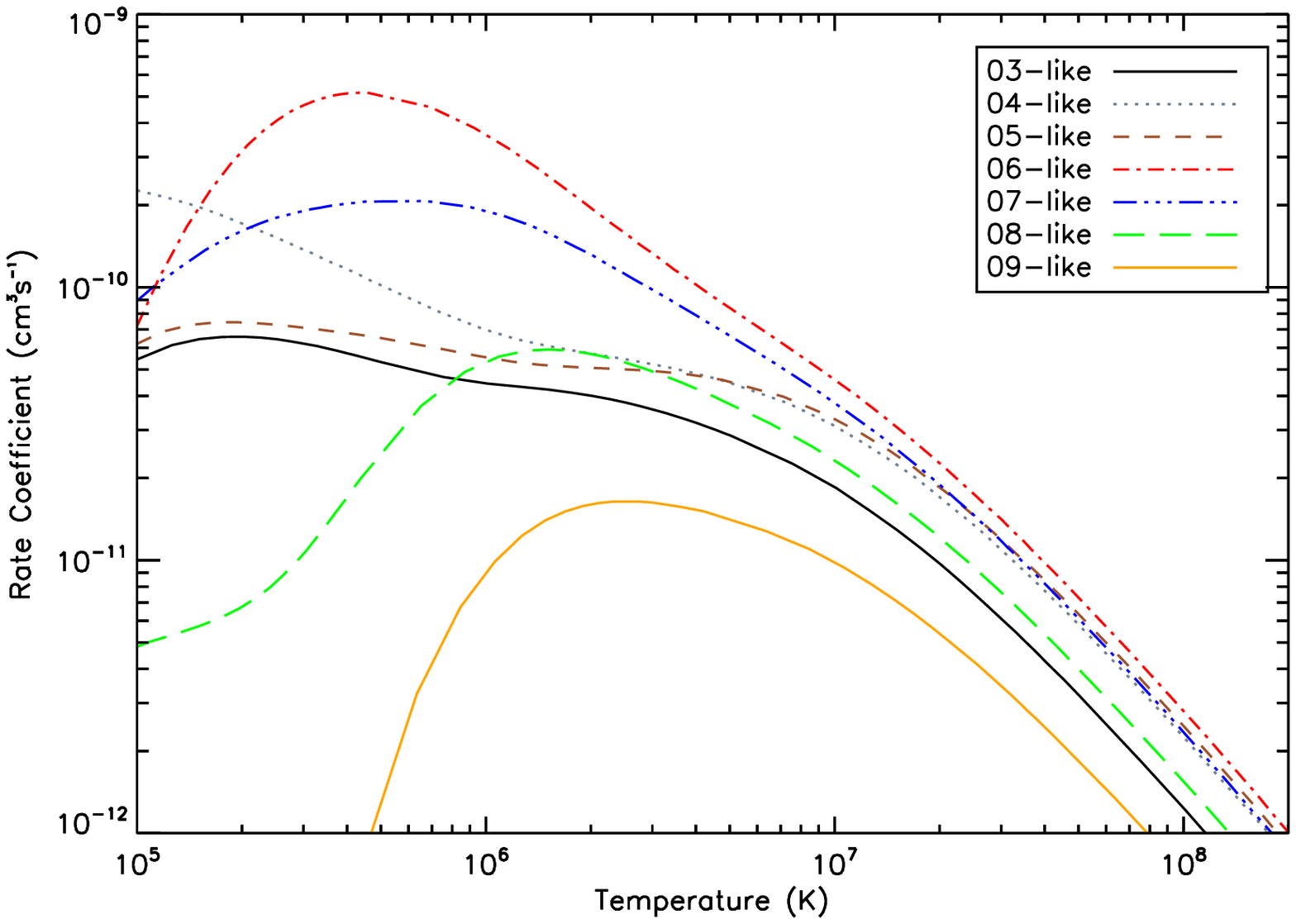}
\caption{DR rate coefficients for 2-2, $\Delta{n}=0$ core-excitation for 03- to 09-like.}
\label{fig:lshelld022}
\end{centering}
\end{figure}

\begin{figure}
\begin{centering}
\includegraphics[width=120mm]{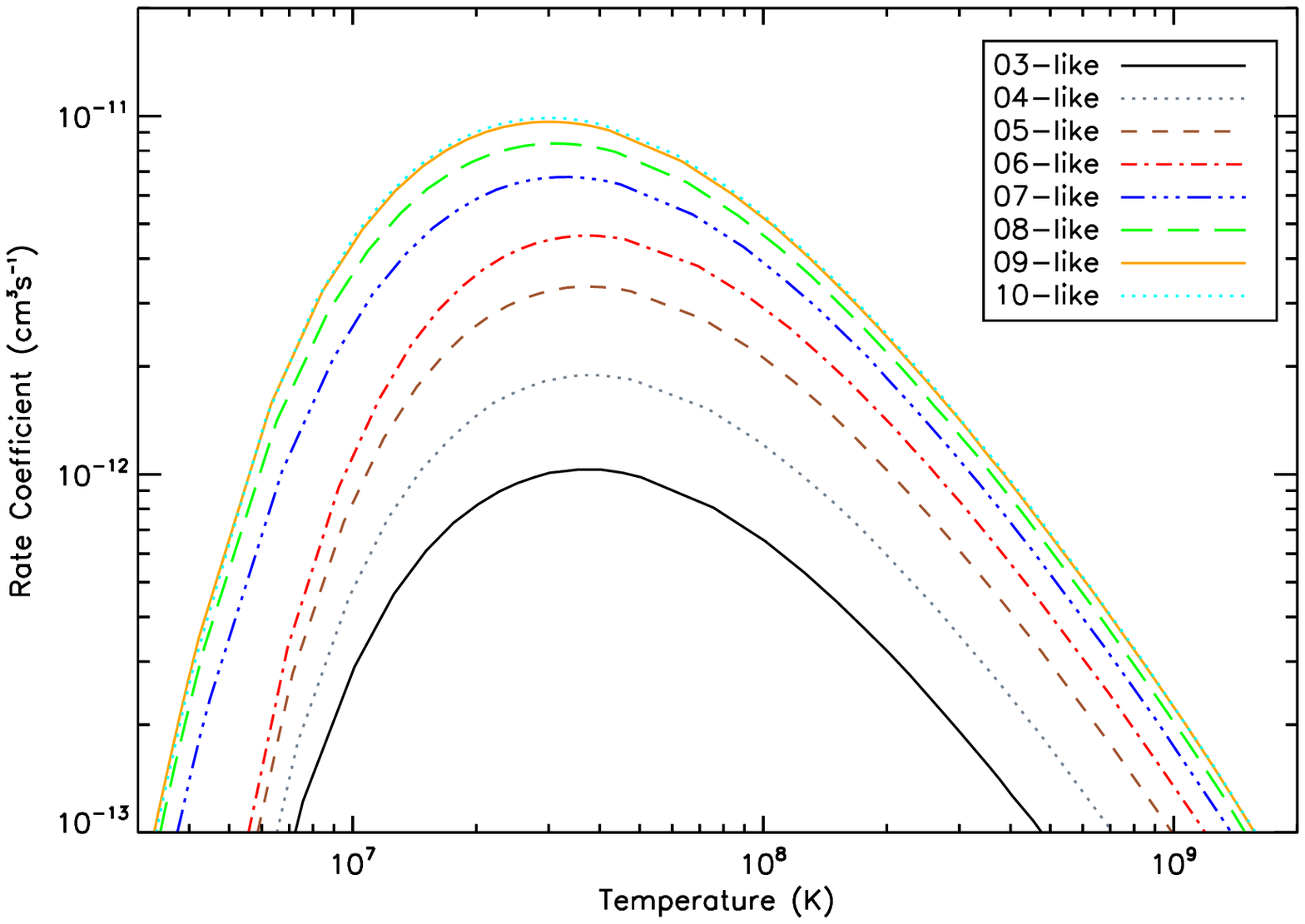}
\caption{DR rate coefficients for 2-3, $\Delta{n}=1$ core-excitation for 03- to 10-like.}
\label{fig:lshelld123}
\end{centering}
\end{figure}

In Figure~\ref{fig:lilikeicr} we have plotted the DR and RR rate coefficients for 03-like
in a similar manner to Figure~\ref{fig:hlikeicr}. The RR rate coefficient drops by a
factor of two due to the K-shell being closed while dominant (for DR) contributions
arise from the 2-2 and 2-3 core-excitations. Nevertheless, RR still contributes
$\sim 60-90$\% of the total recombination rate coefficient around the temperature of
peak abundance. As the L-shell fills, the total RR rate coefficient decreases due to
decreasing L-shell vacancy (and charge somewhat) while the DR increases correspondingly
due to the increasing number of electrons available to be promoted. The two become 
comparable at 10-like (see Figure~\ref{fig:nelikeicr}) when the RR can only start to
fill the M-shell. In Figures \ref{fig:lshelld022} and 
\ref{fig:lshelld123} we have plotted the DR rate coefficients for the 2-2 and 2-3 core 
excitations respectively, with the former covering 03- to 09-like and the latter covering 
03- to 10-like. The 2-2 core-excitation provides the largest contribution to the total DR
rate coefficients when filling the $2s$ shell. After the $2p$ subshell is half filled (06-like), 
the 2-2 DR rate coefficient decreases gradually, being overtaken by the 2-3 core-excitation.
The 2-4 core-excitation provides only a small contribution in 03- and 04-like ($<1$\% at
peak abundance), and was hence neglected from 05-like onwards. In 10-like, the 2-4 core 
excitation was re-introduced as a consistency check now that the 2-2 is closed, however, 
it still provides a minimal contribution of $\sim$5\% around peak abundance.

\subsection{M-shell DR}
\begin{figure}
\begin{centering}
\includegraphics[width=120mm]{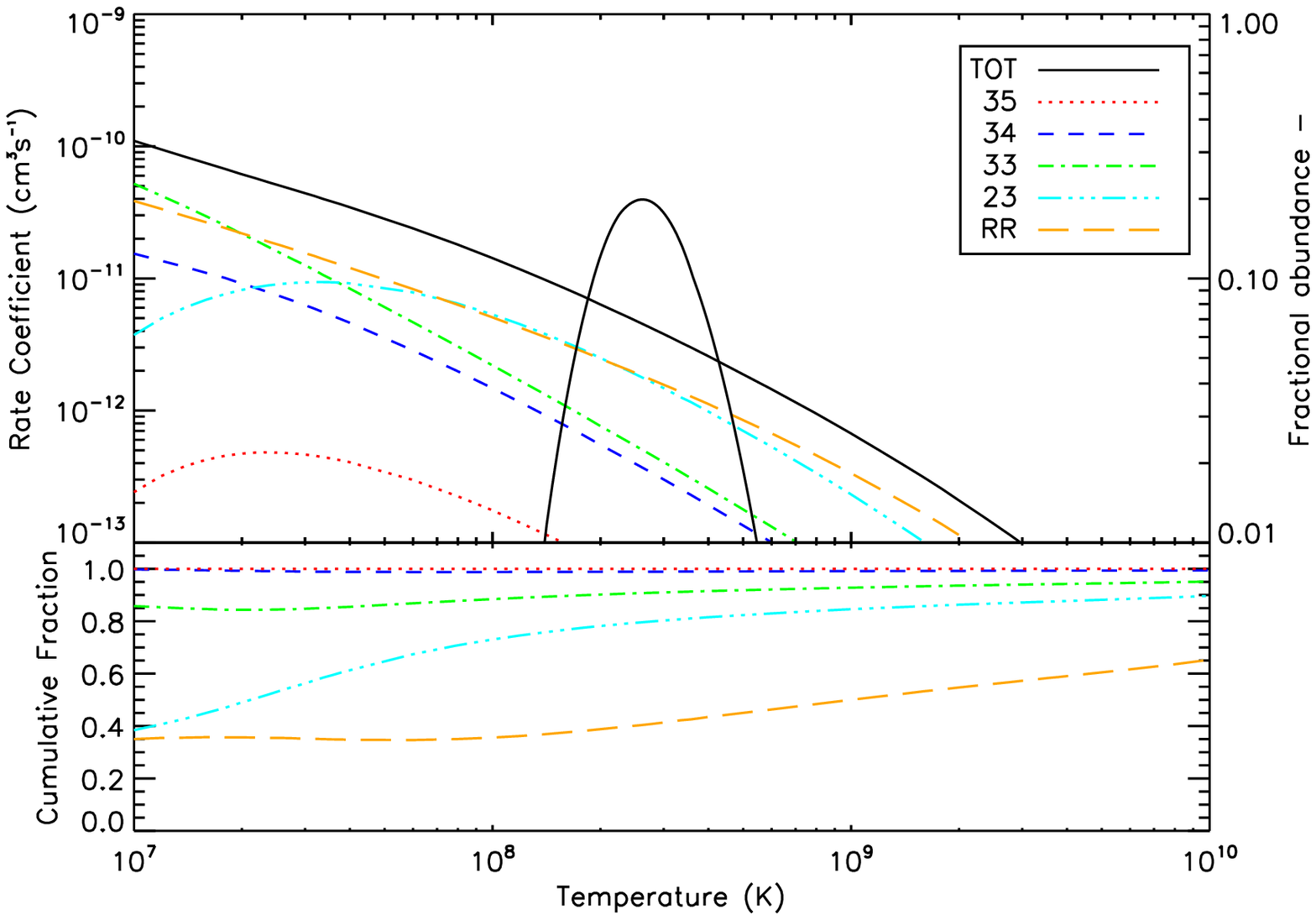}
\caption{As Figure~\ref{fig:hlikeicr}, but for 11-like with core-excitations 2-3, 3-3, 3-4, and 3-5.}
\label{fig:nalikeicr}
\end{centering}
\end{figure}

\begin{figure}
\begin{centering}
\includegraphics[width=120mm]{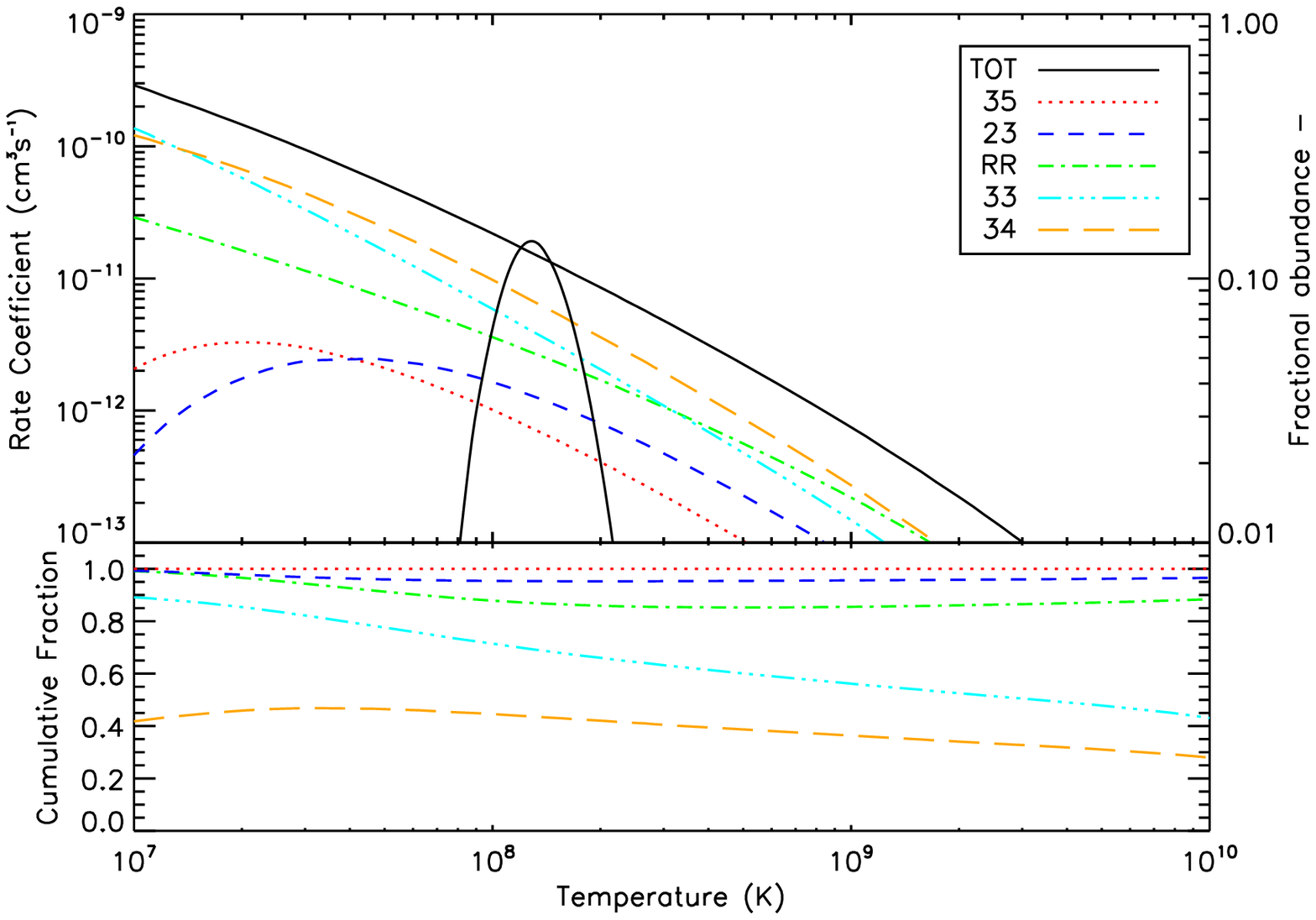}
\caption{As Figure~\ref{fig:hlikeicr}, but for 18-like with core-excitations 2-3, 3-3, 3-4, and 3-5.}
\label{fig:arlikeicr}
\end{centering}
\end{figure}

\begin{figure}
\begin{centering}
\includegraphics[width=120mm]{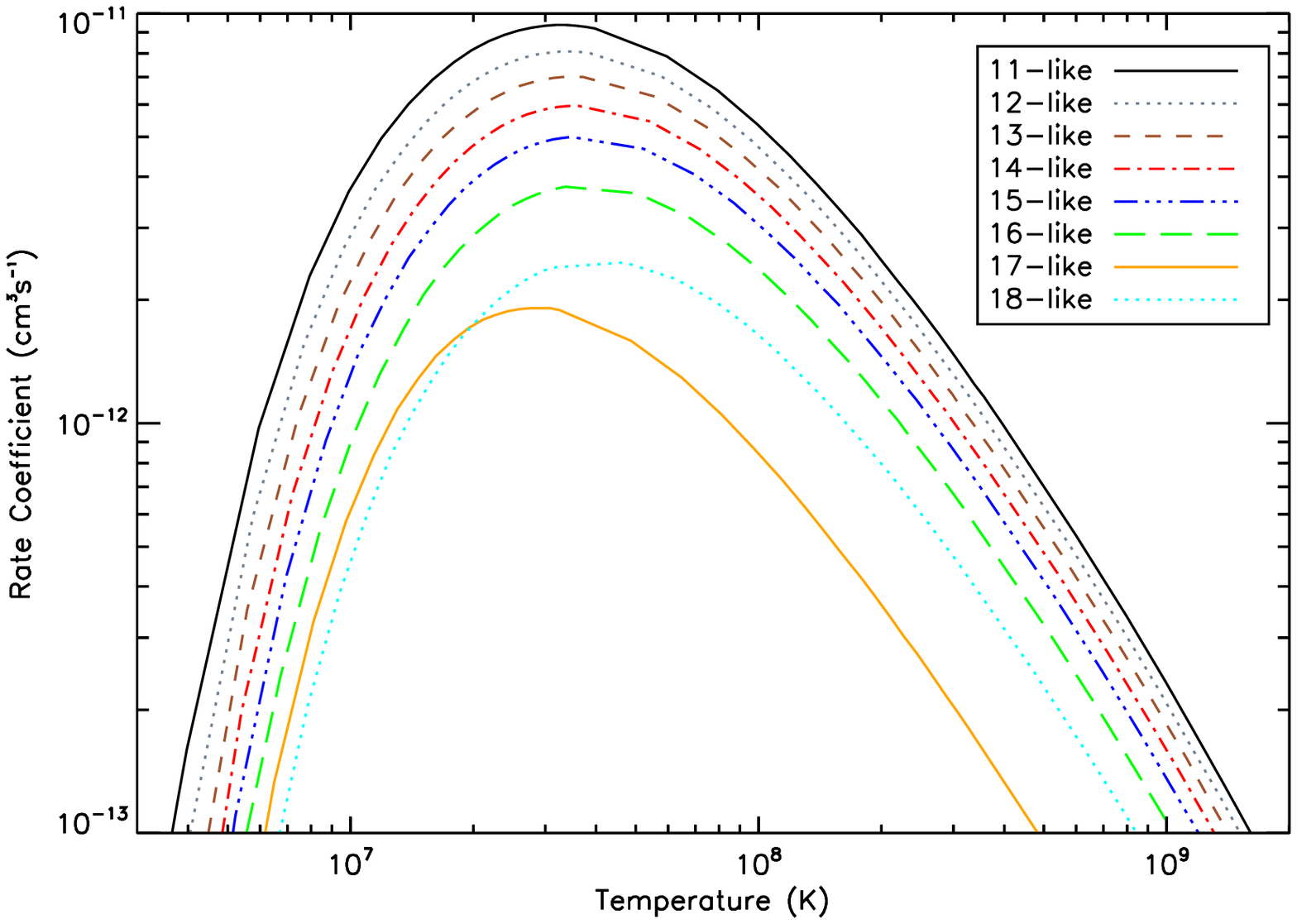}
\caption{DR rate coefficients for 2-3, $\Delta{n}=1$ core-excitation for 11- to 18-like.}
\label{fig:mshelld123}
\end{centering}
\end{figure}

A temperature of 26keV ($3\times{10}^{8}$K) corresponds to the peak abundance of 10-like W. 
Higher charge states will exist, with increasingly small fractional abundance, but may be seen 
spectroscopically. The M-shell is perhaps the deepest shell
in tungsten that ITER will be able to access routinely. The M-shell is also the regime in which RR 
increasingly gives way to DR, contributing $\sim$40\% of the total recombination rate 
coefficient in 11-like, and decreasing to $\sim$15\% in 18-like, around the temperature
of peak abundance (see Figures \ref{fig:nalikeicr} and \ref{fig:arlikeicr}, 
respectively). The inner-shell 2-3 core-excitation provides the largest contribution 
to the total DR rate coefficient in 11-like ($\sim{40}$\%), however, this is quickly 
overtaken by the $\Delta{n}$=0 and outer shell $\Delta{n}$=1 core-excitations of 3-3 and 
3-4, respectively. Again, this can be understood in terms of a simple occupancy/vacancy 
argument. In addition, the 2-3 is increasingly suppressed by core re-arrangement 
autoionizing transitions, viz. an M-shell electron drops down into the L-shell and
ejects another M-shell electron. This process is independent of the Rydberg-$n$, unlike
the initial dielectronic capture. The reduction of the 2-3 core-excitation DR with 
increasing M-shell occupation is shown in Figure~\ref{fig:mshelld123}, where we have
plotted the 2-3 DR rate coefficients for 11-like to 18-like. 

\begin{figure}
\begin{centering}
\includegraphics[width=120mm]{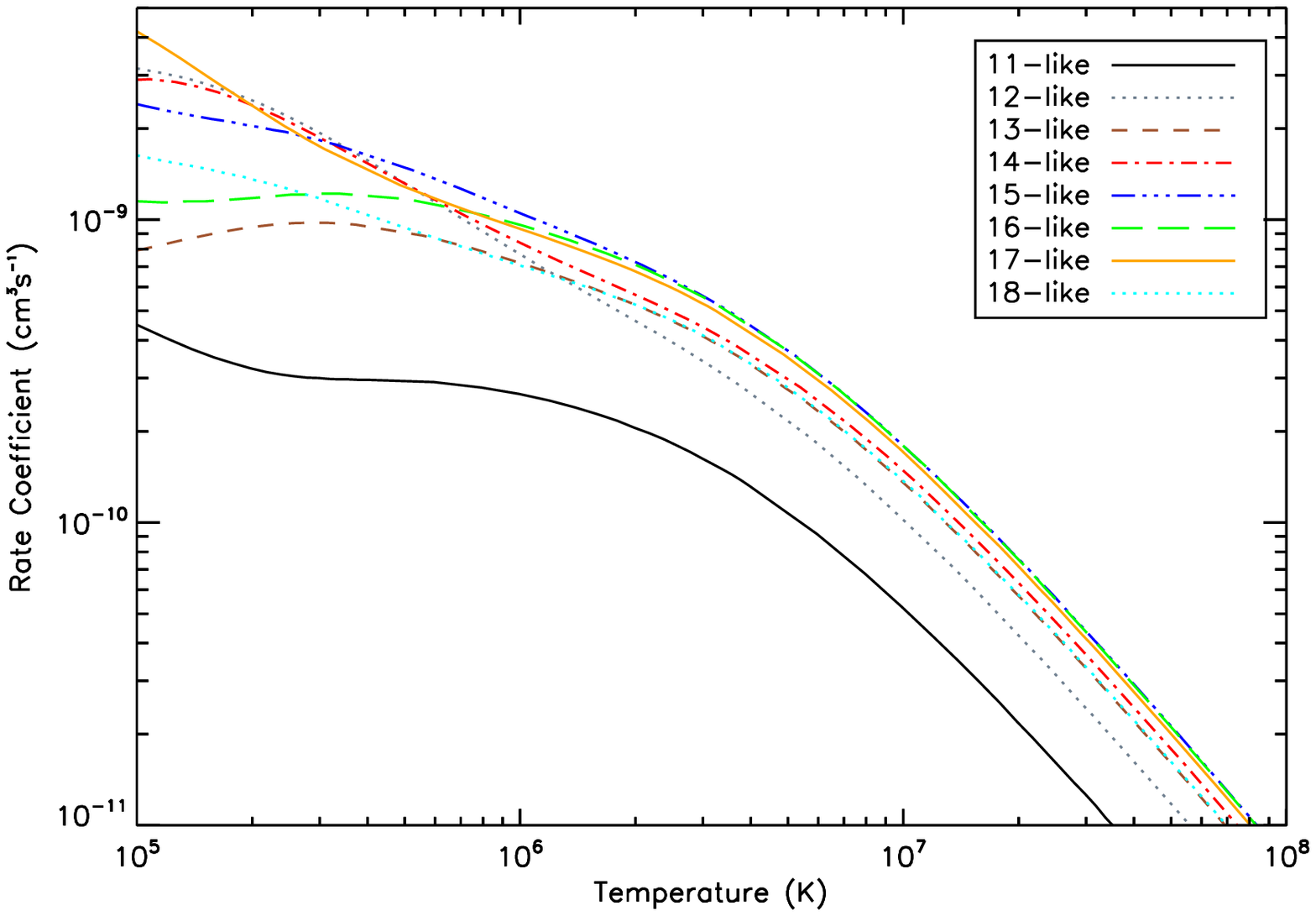}
\caption{DR rate coefficients for 3-3, $\Delta{n}=0$ core-excitation for 11- to 18-like.}
\label{fig:mshelld033}
\end{centering}
\end{figure}

\begin{figure}
\begin{centering}
\includegraphics[width=120mm]{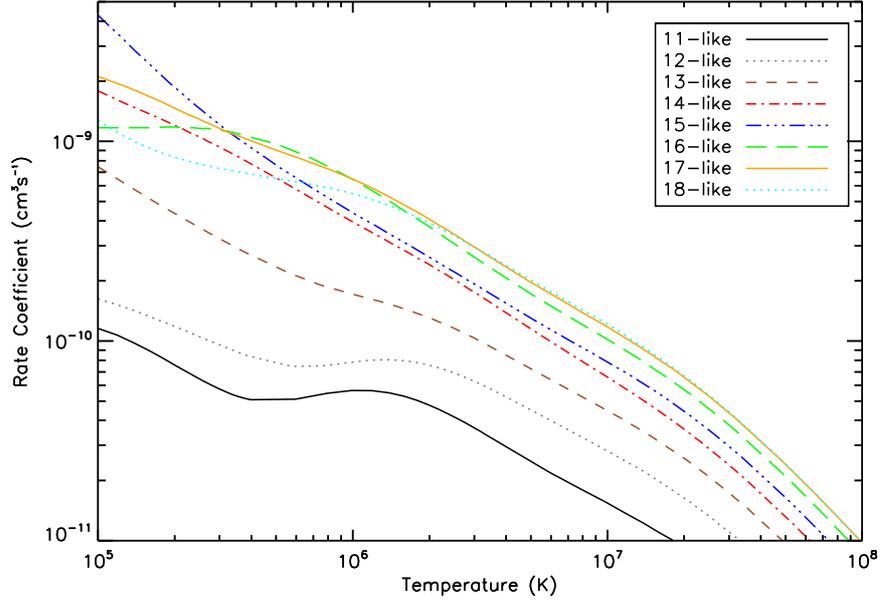}
\caption{DR rate coefficients for 3-4, $\Delta{n}=1$ 3-4 core-excitation for 11- to 18-like.}
\label{fig:mshelld134}
\end{centering}
\end{figure}

The outer-shell $\Delta{n}$=0 (3-3) and $\Delta{n}$=1 (3-4) core-excitations provide the 
largest contributions to the total recombination rate coefficients from 12-like onwards. 
In Figure~\ref{fig:mshelld033}, we have plotted the ($3p^q$) 3-3 core-excitations for 11-like to
18-like, where there is competition between $3p$ occupancy and $3p$ vacancy. It can be seen 
that the 3-3 contribution grows steadily up to 15 and 16-like reaching a maximum value there. 
The rate coefficient then begins to decrease for 17- and 18-like as the $3p$ shell closes, 
leaving only $3d$ vacancies. In Figure~\ref{fig:mshelld134} we have plotted the 3-4 DR rate
coefficients for 11-like to 18-like. The 3-4 rate coefficients increase simply with 
increasing $3p$ occupancy. 

The $\Delta{n}$=2 (3-5) core-excitation again provides only a small contribution
throughout 11- to 18-like. This contribution is at its smallest for 11-like (Figure~
\ref{fig:nalikeicr}), contributing $\sim{1}$\% to the total recombination rate 
coefficient. As with 3-4, the 3-5 DR rate coefficient increases up to 17- and 
18-like, but still only contributes $\sim{5}$\% for the final ion. Despite the small 
contribution, we opted to keep the 3-5 core-excitation as the 2-3 one decreases rapidly with the
filling of the $3p$ shell.

\subsection{Relativistic Configuration Mixing in DR}
\begin{figure}
\begin{centering}
\includegraphics[width=120mm]{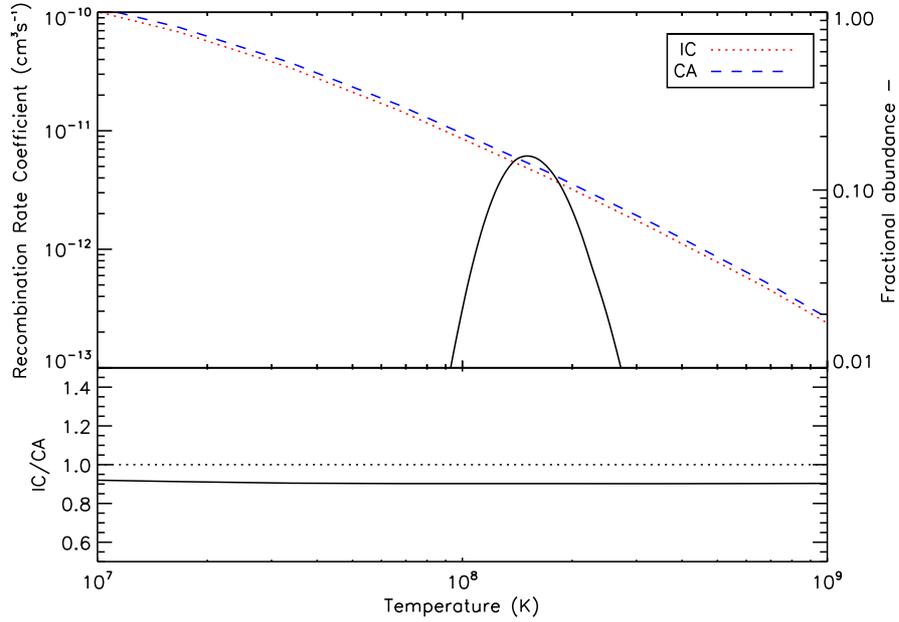}
\caption{Total DR rate coefficients for 16-like 3-4. The bottom plot shows the ratio 
of the IC coefficients to CA. The dotted line indicates a ratio of unity.}
\label{fig:16like34icca}
\end{centering}
\end{figure}

\begin{figure}
\begin{centering}
\includegraphics[width=120mm]{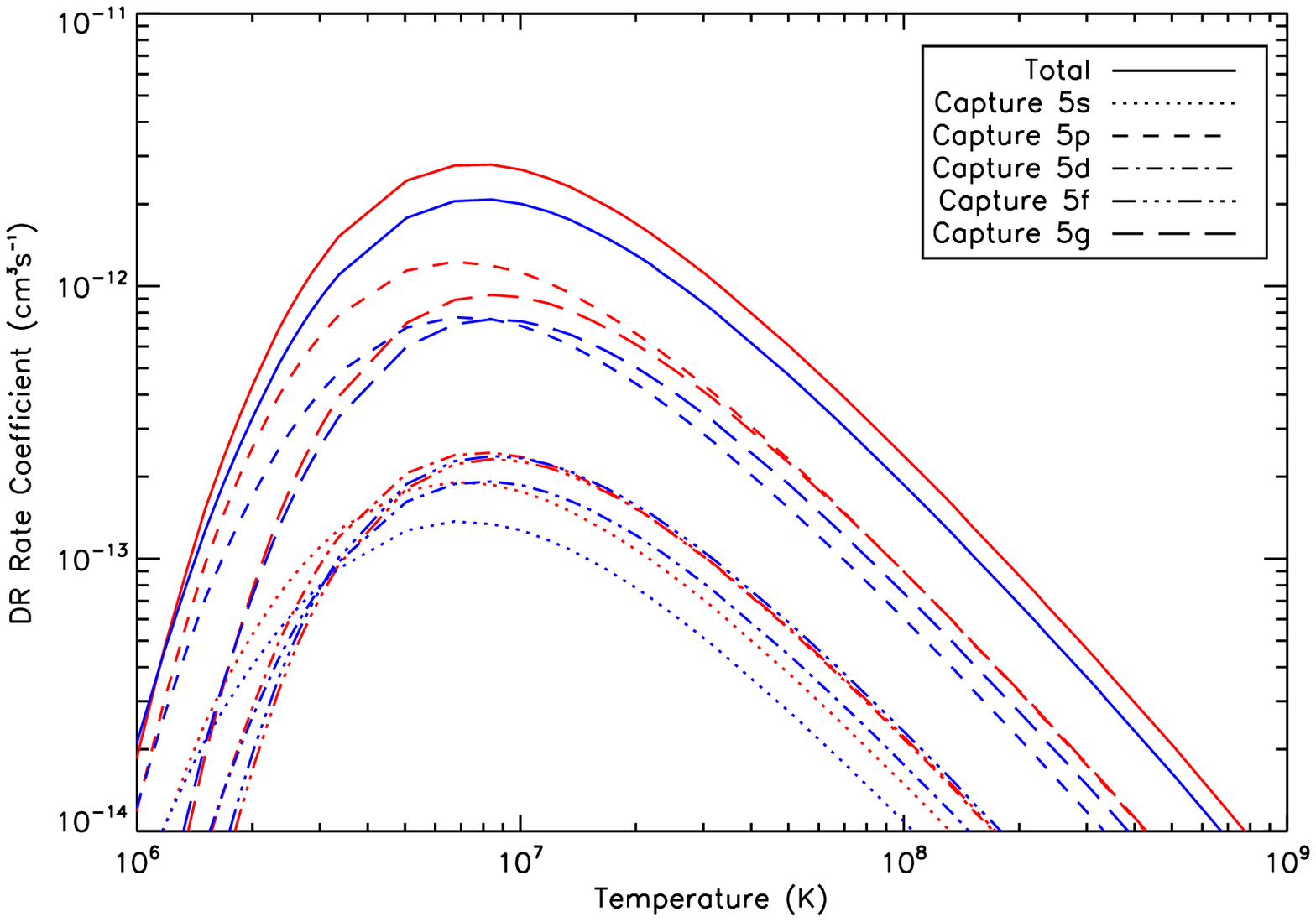}
\caption{Partial DR rate coefficients for 16-like 3-4 capture to $n=5$ for subshells 
$5s-5g$. The red curves correspond to the CA calculation, while the blue correspond to IC.}
\label{fig:16like34partial}
\end{centering}
\end{figure}

Comparing total DR rate coefficients, although convenient, can be somewhat misleading since
non-relativistic configuration mixing and relativistic (e.g. spin-orbit) mixing are
described by unitary transformations of the initial basis wavefunctions. For example, 
in Figure~\ref{fig:16like34icca} we show the total DR rate coefficients for
the 16-like 3-4 core-excitation calculated in IC and CA. It can be seen the agreement between IC and
CA is very good, being $\sim{10}$\% around the temperature of peak abundance. Now, if we consider a 
set of partial DR rate  coefficients for 16-like 3-4, we can see the agreement between IC and CA is
much worse. In Figure~\ref{fig:16like34partial}
we have plotted the partial DR rate coefficients for 16-like 3-4, capture to $n=5$. The best agreement
is for recombination into the $5f$, with IC and CA differing by $\sim{5}$\%. The worst agreement is seen
for recombination into $5p$, where the IC and CA rate coefficients differ by $\sim{33}$\% at peak abundance.
Agreement is no better for $5s$, $5d$, and $5g$, where the IC and CA rate coefficients differ by $\sim{28}$\%,
$\sim{19}$\%, and $\sim{18}$\% respectively. 

\begin{figure}
\begin{centering}
\includegraphics[width=120mm]{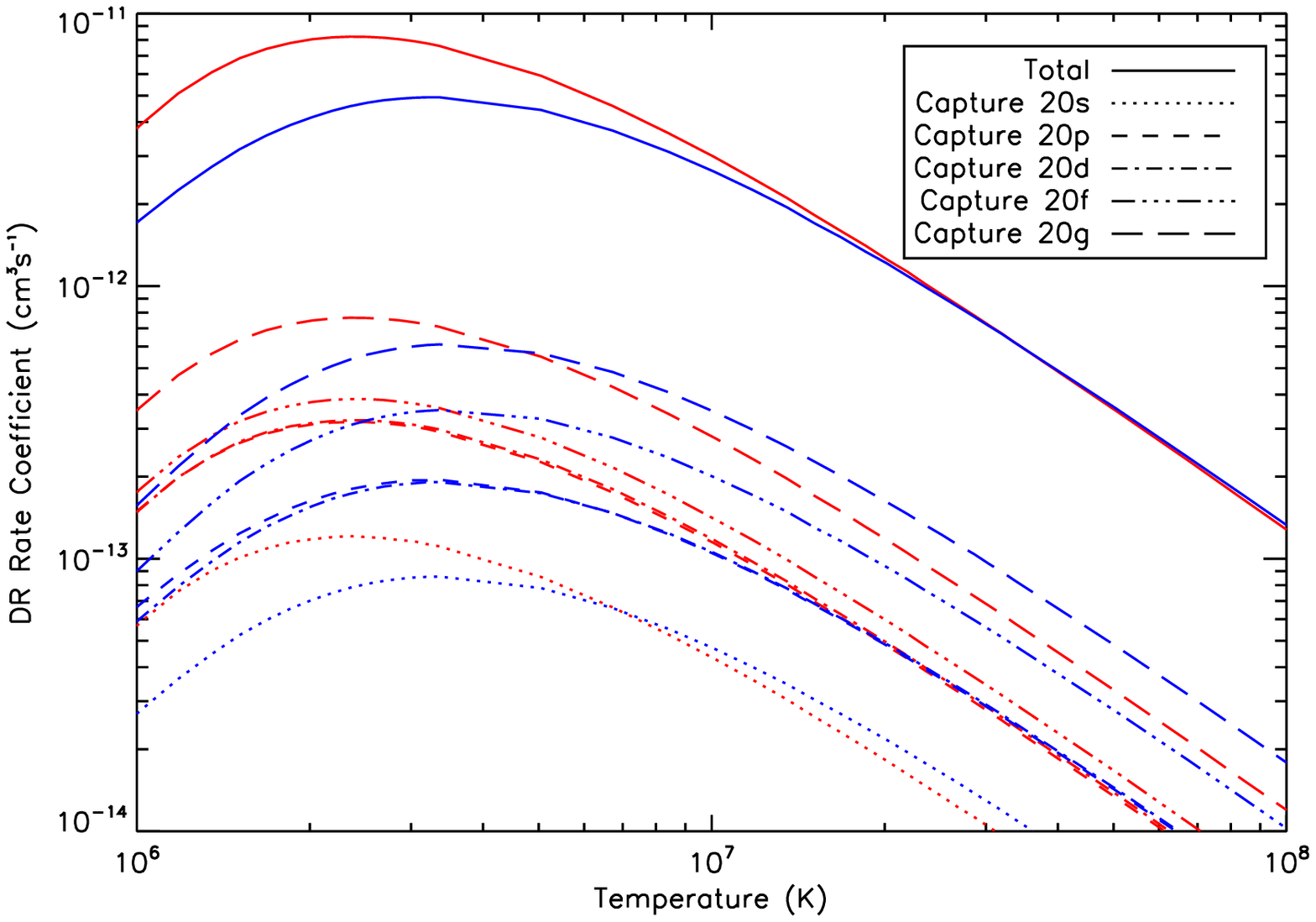}
\caption{Partial DR rate coefficients for 16-like 3-3 capture to $n=20$ for subshells 
$20s-20g$. The red curves correspond to the CA calculation, while the blue correspond to IC.}
\label{fig:16like33partial}
\end{centering}
\end{figure}

The disagreement between partial DR rate coefficients calculated in IC and CA is even more apparent when considering 
the 3-3 core-excitation. In Figure \ref{fig:16like33partial} we have plotted the partial DR rate coefficients for
16-like 3-3, capture to $n=20$. The best agreement occurs for $20p$ and $20d$, where the partials differ by $<10$\% at 
peak abundance. The same cannot be said for $20s$, $20f$, and $20g$, where the IC and CA results differ by $\sim{30}$\%, 
$\sim{72}$\%, and $\sim{51}$\% respectively. These differences highlight the importance of relativistic
mixing for a heavy atom such as tungsten. This effect is not confined to 16-like, and occurs for all ions 
considered in this work. Its subsequent propagation through collisional-radiative modelling
is a topic for future study.

\subsection{Comparison with other DR work}
With the exception of closed-shell ions, not much DR rate coefficient data has been calculated 
for the ions W$^{73+}$ to W$^{56+}$. As ITER will have an operating temperature of up to $\sim{26}$keV 
($\sim{3}\times{10}^8$K), the reactor will be able to access to about 10-like W$^{64+}$. In Tables 
\ref{table:beharcomp} and \ref{table:safronovacomp} we compare our total DR rate coefficients 
for 10-like with those of Behar~\etal \cite{behar99a} and Safronova~\etal \cite{usafronova09b}
respectively, both of whom used the {\sc hullac} \cite{barshalom01a} code. Agreement with the results of
 Behar~\etal is generally good, being $\sim{10}$\% near peak abundance, while low temperatures
illustrate the characteristic sensitivity of DR rate coefficients to near-threshold resonances. However, a significant
difference is noted between these two sets of results and those of Safronova~\etal, where our DR rate
coefficients are larger by $\sim$50\% for temperatures $>2\times{10}^{8}$K. The origin of this 
difference is currently unknown.
In Table \ref{table:pelegcomp} we compare our 18-like total DR rate coefficients with the {\sc hullac} ones of 
Peleg~\etal \cite{peleg98a}. Agreement is better in this case over a wider range of temperatures, 
being $\lesssim$10\% at peak abundance.

\subsection{RR}
\begin{figure}
\begin{centering}
\includegraphics[width=120mm]{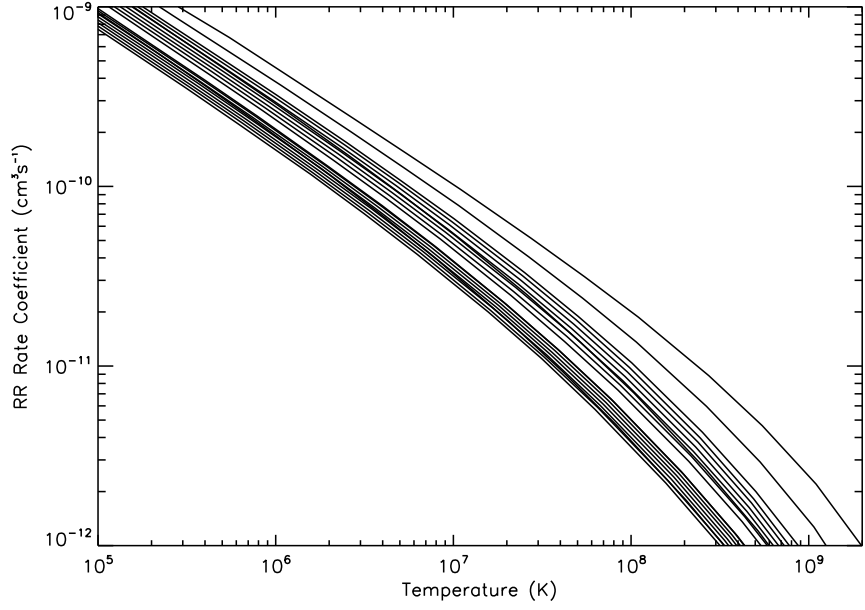}
\caption{Total RR rate coefficients for 00- to 18-like. The top curve is 00-like, 
and the curves below it are 01-like down to 18-like. All this work.}
\label{fig:rrplotsicr}
\end{centering}
\end{figure}

In Figure~\ref{fig:rrplotsicr} we show our total RR rate coefficients from 00-like to
18-like calculated in IC. These include all multipoles up to E40/M39 and
the J\"{u}ttner relativistic correction. The pattern of curves seen corresponds
to the filling of the K-shell and then the L-/M-shell boundary, as noted above. 
As mentioned previously, Trzhaskovskaya~\etal \cite{trzh10a,trzh13a,trzh14a,trzh14b}
have calculated an extensive set of partial and `total' (summed to $n=20$, $\ell=19$) RR rate
coefficients for the whole tungsten isonuclear sequence. Their calculations were fully relativistic, extending to 
$n=20$, $\ell=19$. Comparatively, our {\sc autostructure} calculations extend to $n=999$ 
and $\ell=150$, where values up to $\ell=10$ were treated relativistically in the kappa-averaged approximation.
A non-relativistic dipole top-up was then used to cover the remaining $\ell$ values which become important at low
(non-relativistic) temperatures. In Table \ref{table:trzcomp} we compare the RR rate coefficients of 
Trzhaskovskaya~\etal \cite{trzh10a} for 00-like (fully stripped) to ours over log $T\mathrm{(K)}$ of 3.0 to 10.0.
In this table, we have given our rate coefficients when summed up to $n=999$ and $\ell=150$, as well as the rate 
coefficients when summed up to $n=20$ and $\ell=19$. 
In the case where we do not truncate $n$ and $\ell$ we see very large differences at low temperatures ($>100$\% for 
log $T\mathrm{(K)}\le{3.5}$.) This difference decreases steadily until $\sim{10^9}$K, where it then begins to increase again. 
When  we truncate our $n$ and $\ell$ values to match Trzhaskovskaya~\etal, we find excellent agreement between the two data
sets for log $T\mathrm{(K)}< 9.5$ ($<1$\%). Above log $T\mathrm{(K)}= 9.5$ we note a slight drift away from the results 
of Trzhaskovskaya~\etal, reaching
$\sim{10}$\% at the highest temperature log $T\mathrm{(K)}$=10.0. This is likely due to the use of kappa-average 
wavefunctions by {\sc autostructure}, assuming $<1\%$ accuracy in the results of Trzhaskovskaya~\etal still.
The kappa-average approximation begins to break  down at high temperatures,  or rather at the corresponding high 
electron energies which contribute at such $T$. The underlying photoionization cross sections are falling-off
rapidly in magnitude and such small quatities become increasingly sensitive to
the kappa-average approximation. Such a difference at these temperatures should be of no importance to modelling.
But, it is useful to have a complete set of consistent partial RR data to complement the DR data for collisional-radiative
modelling with ADAS. As already noted, RR is most important for the highest few ionization stages. By 10-like, 
the total DR rate coefficient is comparable to RR  at the temperature of peak abundance. By 18-like, 
the RR rate coefficient contributes only $\approx 10$\% to the total rate coefficient at peak abundance.

\subsection{Comparison with P\"{u}tterich~\etal \& Foster DR+RR}
\begin{figure}
\begin{centering}
\includegraphics[width=120mm]{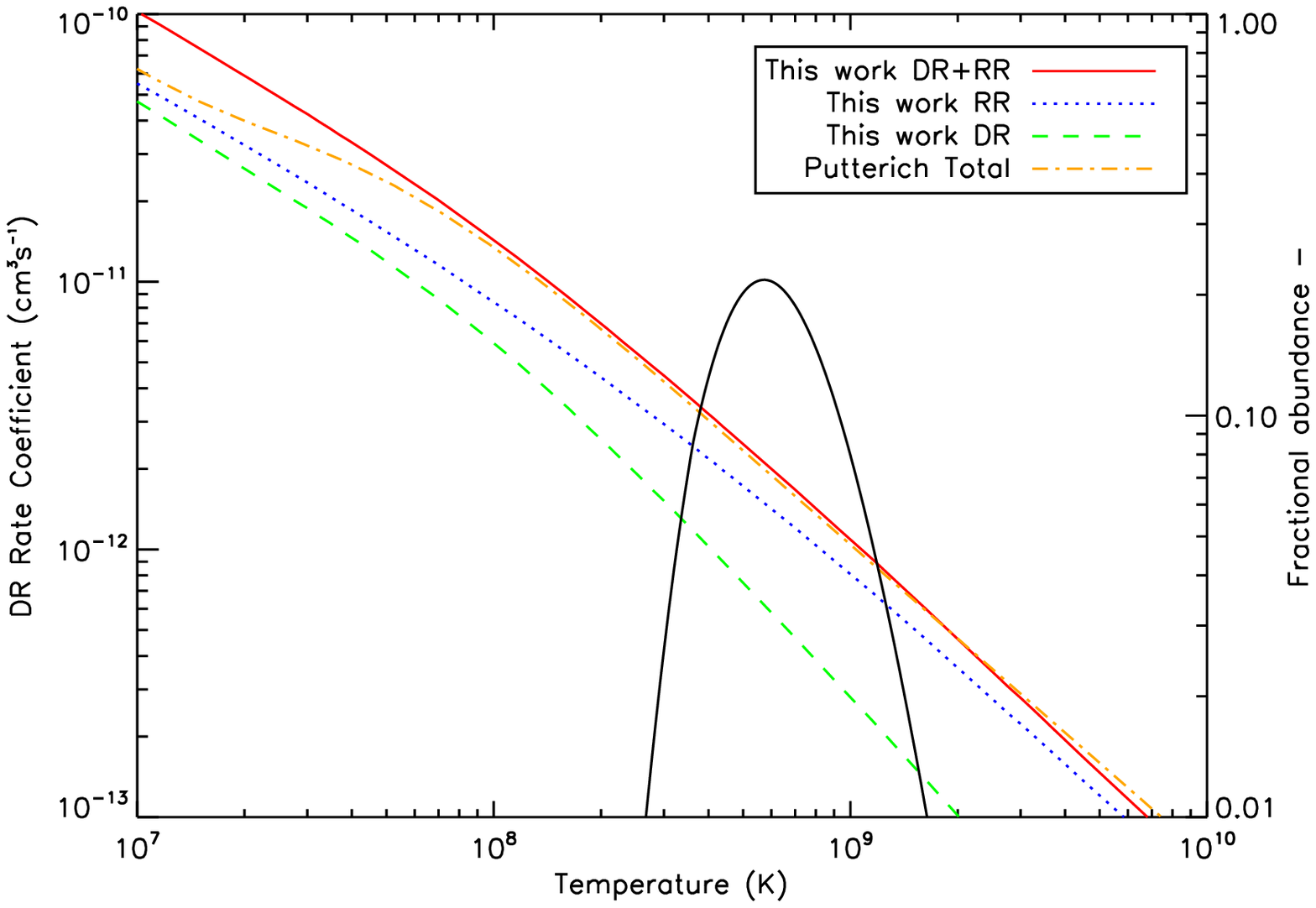}
\caption{Comparison of the present separate DR and RR rate coefficients, together
with their sum total, with the total rate coefficients of P\"{u}tterich~\etal \cite{putterich08a}
 for 06-like.}
\label{fig:puttprevclike}
\end{centering}
\end{figure}

\begin{figure}
\begin{centering}
\includegraphics[width=120mm]{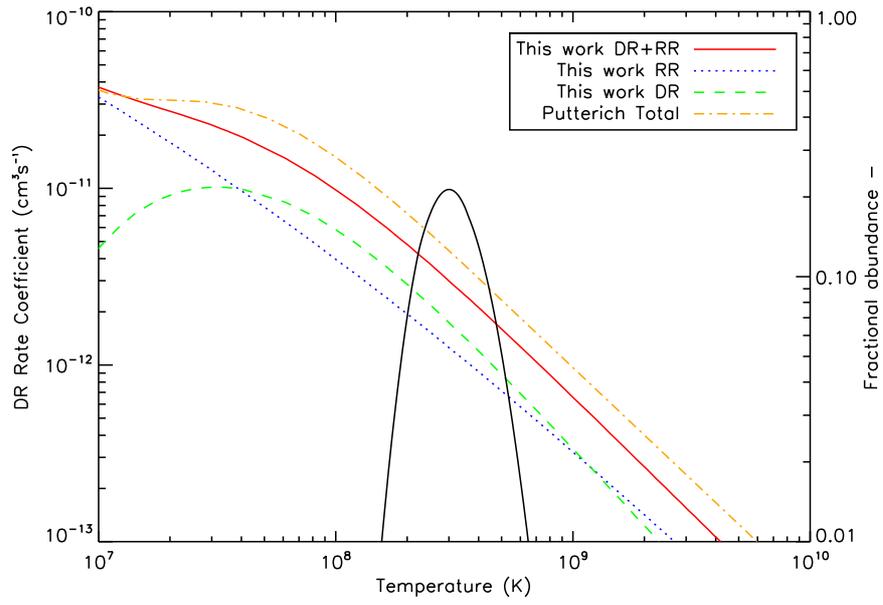}
\caption{As Figure~\ref{fig:puttprevclike}, but for 10-like.}
\label{fig:puttprevnelike}
\end{centering}
\end{figure}

\begin{figure}
\begin{centering}
\includegraphics[width=120mm]{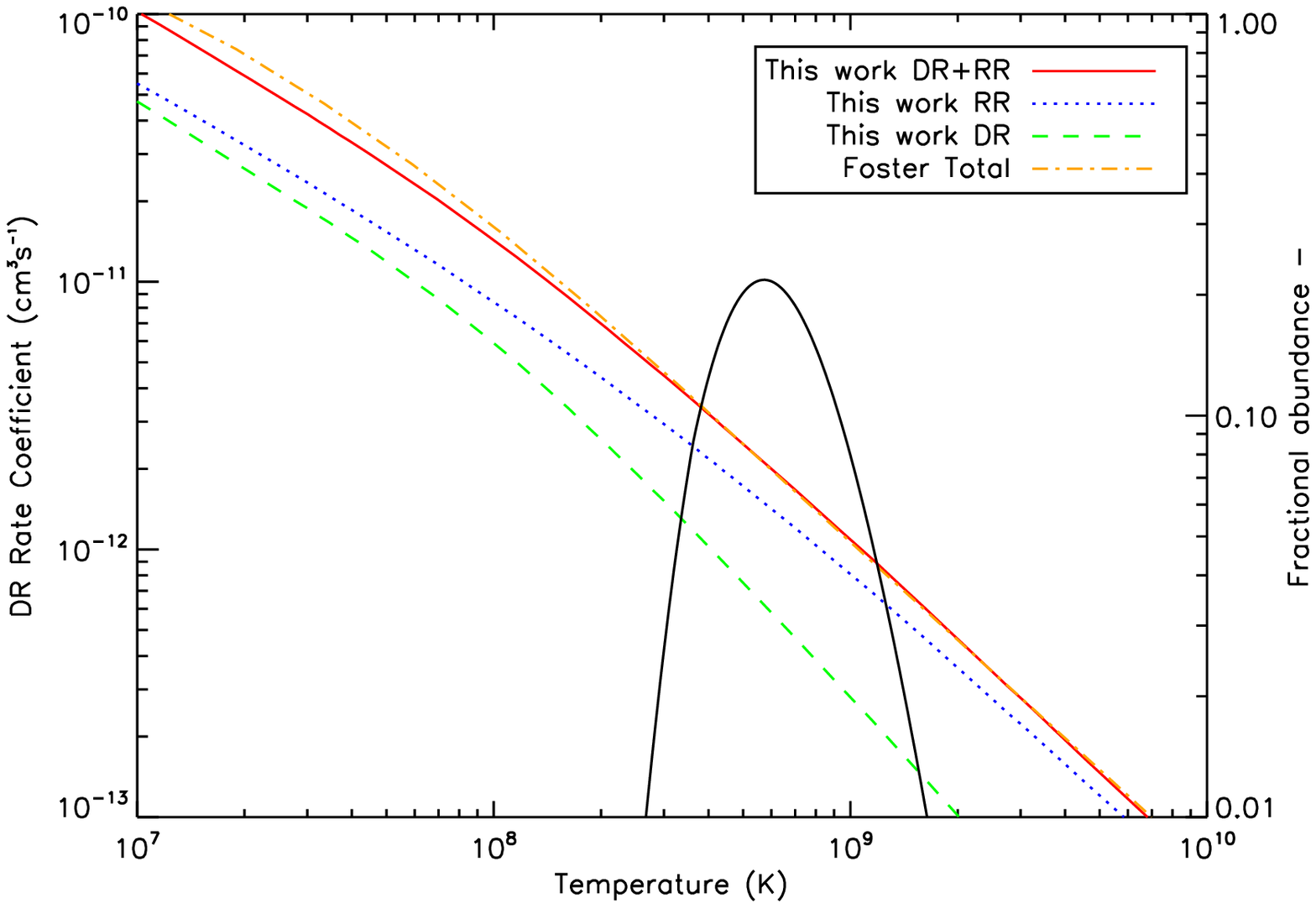}
\caption{Comparison of the present separate DR and RR rate coefficients, together
with their sum total, with the total rate coefficients of Foster \cite{foster08a} for 06-like.}
\label{fig:fostprevclike}
\end{centering}
\end{figure}

\begin{figure}
\begin{centering}
\includegraphics[width=120mm]{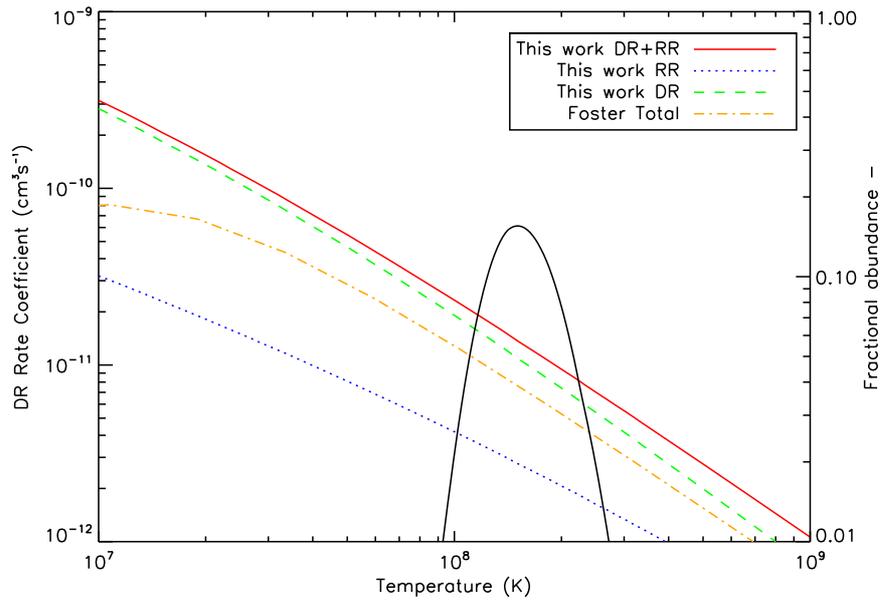}
\caption{Same as Figure~\ref{fig:fostprevclike}, but for 16-like.}
\label{fig:fostprevslike}
\end{centering}
\end{figure}

The P\"{u}tterich~\etal \cite{putterich08a} DR data is ADPAK \cite{post1977a,post1995a}, which uses an
average-atom method, and was further scaled for W$^{22+}$--W$^{55+}$. 
The Foster \cite{foster08a} DR data was calculated using the Burgess General Formula \cite{burgess1965a}.
Both use the same ADAS RR data, which is scaled hydrogenic. 
We now compare our DR+RR results with those P\"{u}tterich~\etal and Foster.
In order to do this, we omit the J\"{u}ttner relativistic correction from our recombination rate coefficients, 
as they did. On comparing our recombination rate coefficients with 
P\"{u}tterich~\etal, we find that there are multiple ions where there is good agreement. For example, in
Figure~\ref{fig:puttprevclike} we have plotted the 06-like
recombination rate coefficients for P\"{u}tterich~\etal, and our DR and RR rate coefficients and their sum.
We find our rate coefficients are in agreement with those of P\"{u}tterich~\etal to $<$10\% at peak abundance.
Some ions are in poor agreement. In Figure~\ref{fig:puttprevnelike} we compare our recombination rate coefficients
with those of  P\"{u}tterich~\etal for 10-like.
The agreement is very poor at peak abundance with a difference of
$>$40\%. For the Foster data, good agreement is again seen in multiple ions. In Figure~\ref{fig:fostprevclike}
we plot our DR, RR, and total recombination rate coefficients along with Foster's total (DR+RR) rate coefficients for 
06-like. The difference between ours and Foster's rate coefficient is even smaller than found with P\"{u}tterich~\etal, 
being $<$1\% at peak abundance. The largest disagreement between ours and Foster's data occurs for 16-like. 
We have plotted ours and Foster's recombination rate coefficients for 16-like in Figure~\ref{fig:fostprevslike}. 
Poor agreement can be seen across a wide temperature range. At peak abundance, ours and Foster's recombination 
rate coefficients differ by $>$40\%. 

The agreement between our present total DR plus RR rate coefficients and those of Foster \cite{foster08a} is 
similar to the agreement between ours and those of P\"{u}tterich~\etal \cite{putterich08a} for 01-like to 
11-like, with the differences being $<$30\% near peak abundance. For 12-like and beyond, 
the P\"{u}tterich~\etal recombination data is in better 
agreement with ours, while Foster's data is consistently smaller than ours. As previously noted, 
DR becomes increasingly important as we move from the L-shell to the M-shell.
Thus, crude/simple methods such as average-atom and the Burgess General Formula can give good
descriptions of DR, but also very poor ones. Also, they are not readily adaptable to delivering the partial 
final-state-resolved data required for collisional-radiative modelling, although the Burgess General Program
underlying his General Formula can do so.

\subsection{Ionization balance}
\begin{figure}
\begin{centering}
\includegraphics[width=120mm]{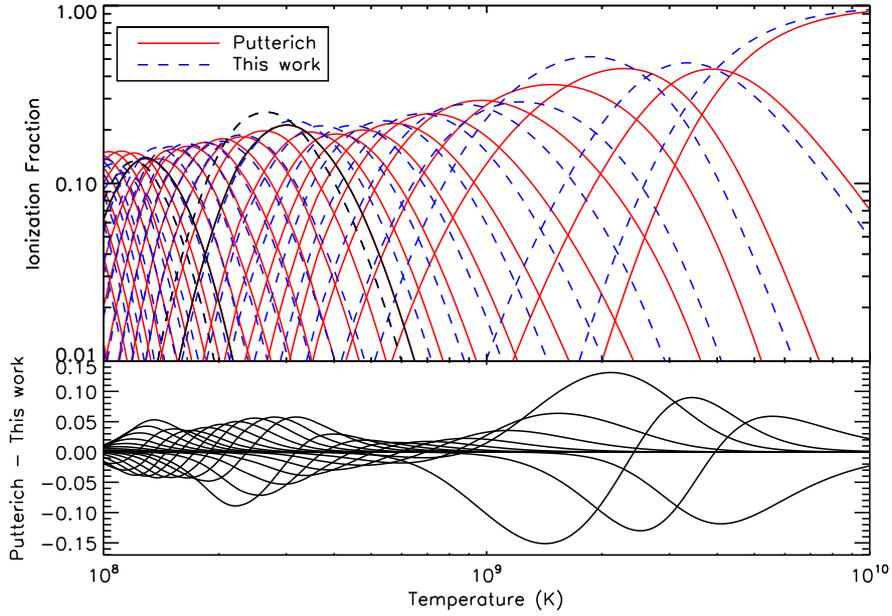}
\caption{Zero-density fractional abundances of tungsten ionization stages, calculated using 
P\"{u}tterich~\etal \cite{putterich08a} recombination data (red, solid curves) and the present
 recombination data (blue, dashed curves). The black curves, from right to left, indicate 10-like 
and 18-like. Both use the ionization rate coefficients from Loch \etal \protect\cite{loch05a}.}
\label{fig:puttprevcom1}
\end{centering}
\end{figure}

\begin{figure}
\begin{centering}
\includegraphics[width=120mm]{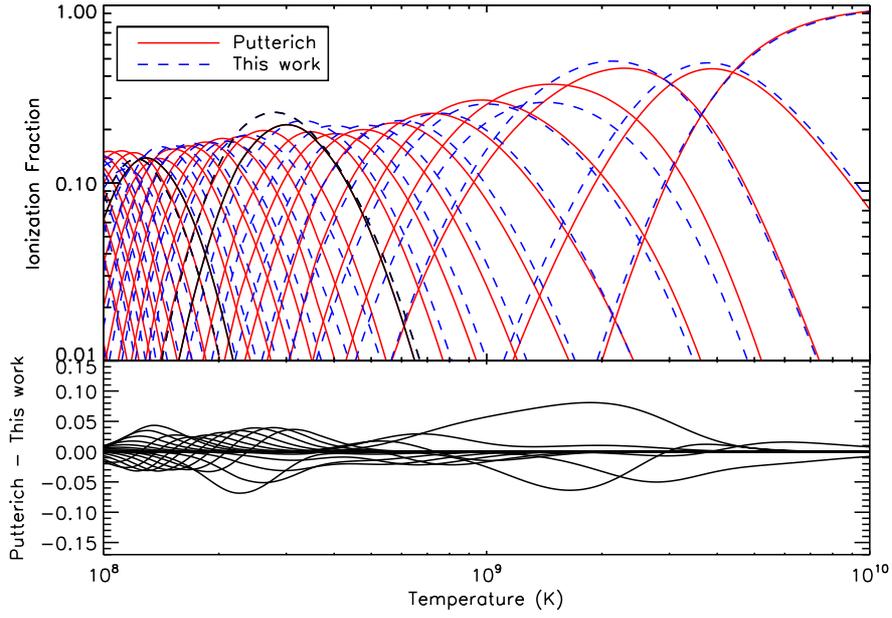}
\caption{As Figure~\protect\ref{fig:puttprevcom1}, but with the J\"{u}ttner correction
removed from our data.}
\label{fig:puttprevcom2}
\end{centering}
\end{figure}

In order to compare the effect of our new recombination data, on the zero-density ionization balance,
with that of  P\"{u}tterich~\etal \cite{putterich08a}, we replaced their recombination data with our 
new DR+RR data for 00-like to 18-like tungsten. In Figure~\ref{fig:puttprevcom1}, we compare the 
ionization balance obtained  with this new data set with the original one of P\"{u}tterich~\etal. 
A large discrepancy is immediately apparent,
namely that our peak abundance fractions have shifted relative to those of P\"{u}tterich~\etal. 
This has a simple explanation, in that our data has the J\"{u}ttner relativistic 
correction applied. By excluding this correction, our ionization fraction moves into 
better agreement with the P\"{u}tterich~\etal fraction, as seen in Figure
\ref{fig:puttprevcom2}.

\begin{figure}
\begin{centering}
\includegraphics[width=120mm]{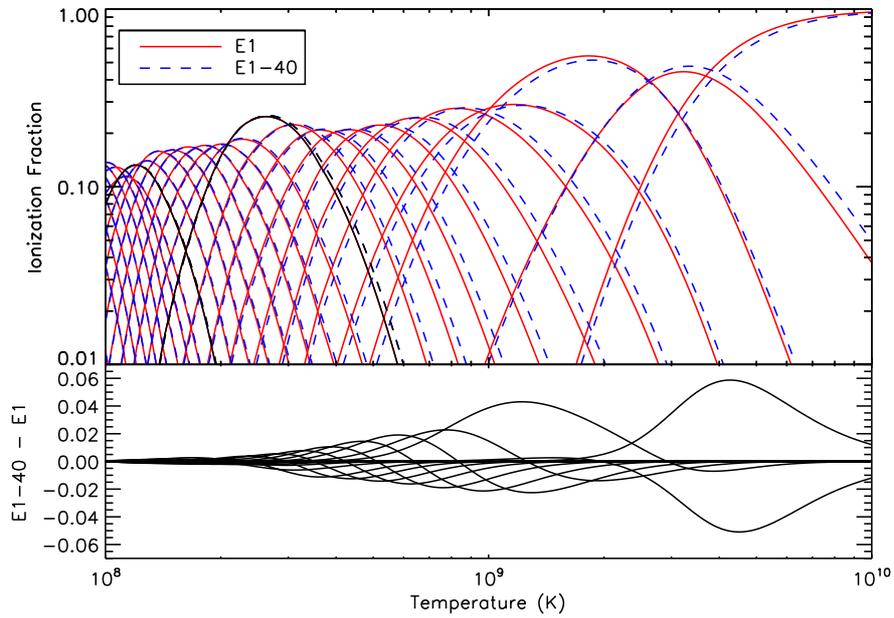}
\caption{Zero-density fractional abundances of tungsten ionization stages, 
calculated using RR rate coefficients with dipole only (red, solid curves)
and E1-40/M39 multipoles included (blue, dashed curves). All this work.}
\label{fig:multipolecom}
\end{centering}
\end{figure}

Electric and magnetic multipole radiation contributions to the RR rate coefficients 
become important at high temperatures \cite{pratt1973a}. In Figure~\ref{fig:multipolecom}
we have plotted the ionization balance using RR rate coefficients where only electric 
dipole radiation is included, and using RR rate coefficients where electric and magnetic 
multipoles up to E40 and M39 have been included. The inclusion of higher multipoles 
increases the peak abundance temperature of the highest-charge ions as expected, however, the
peak abundance temperature only changes by $\sim{4}$\% for 01- and 02-like.
This shift decreases rapidly to zero towards 18-like as DR becomes dominant over RR.

\section{Conclusion}
Large uncertainties exist in the tungsten ionization balance over a wide range of
temperatures (charge-states) found in magnetic fusion plasmas. This ranges from
the cool edge plasma right through to the hot core plasma. The cause is the simplified
treatment of DR, using either the average-atom or Burgess General Formula approaches.
We have embarked on a program of work to address this deficiency.
In this paper, we have reported on the calculation of CA \& IC DR and RR rate 
coefficients for 00-like to 18-like tungsten (W$^{74+}$ to W$^{56+}$ ions) using {\sc autostructure}.
In particular, we retain the partial final-state-resolved coefficients in a suitable
form ({\it adf09} and {\it adf48} files) which are necessary for the collisional-radiative
modelling of tungsten ions at the densities found in magnetic fusion plasmas. 

We have compared our total DR rate coefficients to the results of calculations provided by 
Behar~\etal~\cite{behar99a} and Safronova~\etal~\cite{usafronova09b} for 10-like,
and Peleg~\etal~\cite{peleg98a} for 18-like tungsten. Good agreement is found
between our rate coefficients and those of \cite{behar99a} and \cite{peleg98a}
for 10-like and 18-like, differing by $\sim10$\% at the peak abundance temperature.
Poor agreement was found when comparing with the 10-like results of \cite{usafronova09b}, 
with differences of $\sim$50\%. 

RR dominates the recombination of the highest charge states (K-shell ions) but DR
becomes increasingly important as the L-shell fills and by 10-like it is (just)
larger around the temperatures of peak abundance. For more lowly ionized tungsten, 
beyond 10-like, the importance of RR rapidly diminishes.

We have calculated a new zero-density ionization balance for tungsten by replacing the 
P\"{u}tterich~\etal~\cite{putterich08a} recombination with our new DR+RR data for 00-like to 18-like.
Large differences result, both in the peak abundance temperatures
and the ionization fractions, due largely to our inclusion of the J\"{u}ttner
relativistic correction to the Maxwell-Boltzmann electron distribution. A further, smaller,
difference arises from our inclusion of high electric and magnetic multipole radiation 
which causes a slight shift in the peak abundance temperatures of higher ionization 
stages (in particular, K shell ions). 

This paper has presented the first step in a larger programme of work within {\it The Tungsten Project}.
The next paper will cover DR/RR calculations for 19-like to 36-like tungsten, with the possibility of 
modelling a density-dependent ionization balance. Our ultimate goal within {\it The Tungsten Project}
is to calculate partial and total DR/RR rate coefficients for the entire isoelectronic sequence of tungsten.
This will replace the less reliable data used at present, which is mostly based-on average-atom and
the Burgess General Formula (for DR), and which gives rise to large uncertainties in 
the tungsten ionization balance.

\begin{acknowledgments}
SPP, NRB, and MGOM acknowledge the support of EPSRC grant EP/1021803
to the University of Strathclyde. 
One of us (SPP) would like to thank Stuart Henderson, 
Stuart Loch, and Connor Ballance for useful discussions.

\end{acknowledgments}

\newpage
\bibliography{simonpreval}
\newpage

\begin{table}[H]
\caption{Core excitations included in the DR rate coefficient calculations for  W$^{73+}$ to W$^{56+}$.
\newline
All core-excitations have been calculated in IC and CA.}
\begin{tabular}{@{}llllll}
\hline
Ion-like & Symbol & Core excitations & Ion & Symbol & Core excitations \\
\hline
01-like   & W$^{73+}$  &  1-2, 1-3            & 10-like  & W$^{64+}$  &  2-3, 2-4           \\  
02-like   & W$^{72+}$  &  1-2, 1-3            & 11-like  & W$^{63+}$  &  2-3, 3-3, 3-4, 3-5 \\
03-like   & W$^{71+}$  &  1-2, 2-2, 2-3, 2-4  & 12-like  & W$^{62+}$  &  2-3, 3-3, 3-4, 3-5 \\
04-like   & W$^{70+}$  &  1-2, 2-2, 2-3, 2-4  & 13-like  & W$^{61+}$  &  2-3, 3-3, 3-4, 3-5 \\
05-like   & W$^{69+}$  &  2-2, 2-3            & 14-like  & W$^{60+}$  &  2-3, 3-3, 3-4, 3-5 \\
06-like   & W$^{68+}$  &  2-2, 2-3            & 15-like  & W$^{59+}$  &  2-3, 3-3, 3-4, 3-5 \\
07-like   & W$^{67+}$  &  2-2, 2-3            & 16-like  & W$^{58+}$  &  2-3, 3-3, 3-4, 3-5 \\
08-like   & W$^{66+}$  &  2-2, 2-3            & 17-like  & W$^{57+}$  &  2-3, 3-3, 3-4, 3-5 \\
09-like   & W$^{65+}$  &  2-2, 2-3            & 18-like  & W$^{56+}$  &  2-3, 3-3, 3-4, 3-5 \\
\hline
\label{table:drcorex}
\end{tabular}
\end{table}

\begin{table}[H]
\renewcommand{\arraystretch}{0.8}
\caption{Set of configurations used for the 14-like 3-3 and 3-4 core-excitation calculations.
\newline
Configurations marked with an * were included as mixing configurations.}
\begin{tabular}{@{}lllll}
\hline
3-3 & & 3-4 & \\
$N$-electron & $(N+1)$-electron & $N$-electron & $(N+1)$-electron \\
\hline
$3s^2 3p^2$   & $3s^2 3p^3$    & $3s^2 3p^2$   & $3s^2 3p^2 4s$  & $3s^2 3p 4s^2$  \\
$3s^2 3p 3d$  & $3s^2 3p^2 3d$ & $3s^2 3p 3d$  & $3s^2 3p^2 4p$  & $3s^2 3p 4s 4p$ \\
$3s 3p^3$     & $3s^2 3p 3d2$  & $3s^2 3p 4s$  & $3s^2 3p^2 4d$  & $3s^2 3p 4s 4d$ \\
$3s 3p^2 3d$  & $3s 3p^4$      & $3s^2 3p 4p$  & $3s^2 3p^2 4f$  & $3s^2 3p 4s 4f$ \\
$*3p^4$       & $3s 3p^3 3d$   & $3s^2 3p 4d$  & $3s^2 3p 3d 4s$ & $3s^2 3p 4p^2$  \\
$*3s^2 3d^2$  & $3s 3p^2 3d^2$ & $3s^2 3p 4f$  & $3s^2 3p 3d 4p$ & $3s^2 3p 4p 4d$ \\
              & $*3p^5$        & $3s 3p^3$     & $3s^2 3p 3d 4d$ & $3s^2 3p 4p 4f$ \\
              & $*3p^4 3d$     & $3s 3p^2 3d$  & $3s^2 3p 3d 4f$ & $3s^2 3p 4d^2$  \\
              & $*3s^2 3d^3$   & $3s 3p^2 4s$  & $3s 3p^3 4s $   & $3s^2 3p 4d 4f$ \\
              &                & $3s 3p^2 4p$  & $3s 3p^3 4p$    & $3s^2 3p 4f^2$  \\
              &                & $3s 3p^2 4d$  & $3s 3p^3 4d$    & $3s 3p^2 4s^2$  \\
              &                & $3s 3p^2 4f$  & $3s 3p^3 4f$    & $3s 3p^2 4s 4p$ \\
              &                & $*3p^4$       & $3s 3p^2 3d 4s$ & $3s 3p^2 4s 4d$ \\
              &                & $*3s^2 3d^2$  & $3s 3p^2 3d 4p$ & $3s 3p^2 4s 4f$ \\
              &                & $*3s^2 3d 4s$ & $3s 3p^2 3d 4d$ & $3s 3p^2 4p^2$  \\
              &                & $*3s^2 3d 4p$ & $3s 3p^2 3d 4f$ & $3s 3p^2 4p 4d$ \\
              &                & $*3s^2 3d 4d$ & $*3p^4 4s$      & $3s 3p^2 4p 4f$ \\
              &                & $*3s^2 3d 4f$ & $*3p^4 4p$      & $3s 3p^2 4d^2$  \\
              &                &               & $*3p^4 4d$      & $3s 3p^2 4d 4f$ \\
              &                &               & $*3p^4 4f$      & $3s 3p^2 4f^2$  \\
              &                &               & $*3s^2 3d^2 4s$ & $*3s^2 3d 4s^2$  \\
              &                &               & $*3s^2 3d^2 4p$ & $*3s^2 3d 4s 4p$ \\
              &                &               & $*3s^2 3d^2 4d$ & $*3s^2 3d 4s 4d$ \\
              &                &               & $*3s^2 3d^2 4f$ & $*3s^2 3d 4s 4f$ \\
              &                &               &                 & $*3s^2 3d 4p^2$  \\
              &                &               &                 & $*3s^2 3d 4p 4d$ \\
              &                &               &                 & $*3s^2 3d 4p 4f$ \\
              &                &               &                 & $*3s^2 3d 4d^2$  \\
              &                &               &                 & $*3s^2 3d 4d 4f$ \\
              &                &               &                 & $*3s^2 3d 4f^2$  \\
\hline
\label{table:14likeconf}
\end{tabular}
\renewcommand{\arraystretch}{1.0}
\end{table}

\begin{table}
\renewcommand{\arraystretch}{0.8}
\caption{Comparison of total RR rate coefficients for 00-like between those calculated by 
Trzhaskovskaya~\etal \cite{trzh10a}, this work, and the \% difference between the 
two\footnote{The \% difference is calculated as 
$(\alpha_{\mathrm{DR Present}}-\alpha_{\mathrm{DR Trzhaskovskaya}})/\alpha_{\mathrm{DR Trzhaskovskaya}}$.}.
The `Cut' columns correspond to the total 
RR rate coefficient where we restrict the partial sum up to $n=20$ and $l=19$ so as
to match that of \cite{trzh10a}. Quantities in square brackets are powers of ten, for example, $1.00[-1]=1.00\times{10}^{-1}$}
\begin{tabular}{@{}llllll}
\hline
Log $T$ (K) & Trzhaskovskaya~\etal & This work (No Cut) & This work (Cut) & \%Diff.(No
Cut) & \%Diff.(Cut) \\
\hline
 3.0 & 1.17[-08] & 3.00[-08] & 1.17[-08] &   156 &  0.0 \\
 3.5 & 6.56[-09] & 1.46[-08] & 6.60[-09] &   123 &  0.6 \\ 
 4.0 & 3.69[-09] & 7.28[-09] & 3.71[-09] &  97.3 &  0.5 \\
 4.5 & 2.07[-09] & 3.64[-09] & 2.08[-09] &  75.8 &  0.5 \\
 5.0 & 1.16[-09] & 1.83[-09] & 1.17[-09] &  57.8 &  0.9 \\
 5.5 & 6.45[-10] & 9.09[-10] & 6.48[-10] &  40.9 &  0.5 \\
 6.0 & 3.51[-10] & 4.47[-10] & 3.53[-10] &  27.4 &  0.6 \\
 6.5 & 1.85[-10] & 2.16[-10] & 1.86[-10] &  16.8 &  0.5 \\
 7.0 & 9.30[-11] & 1.02[-10] & 9.35[-11] &   9.7 &  0.5 \\
 7.5 & 4.41[-11] & 4.64[-11] & 4.43[-11] &   5.2 &  0.5 \\
 8.0 & 1.95[-11] & 2.00[-11] & 1.95[-11] &   2.6 &  0.0 \\
 8.5 & 7.71[-12] & 7.80[-12] & 7.69[-12] &   1.2 & -0.3 \\
 9.0 & 2.46[-12] & 2.46[-12] & 2.44[-12] &   0.0 & -0.8 \\
 9.5 & 5.42[-13] & 5.26[-13] & 5.24[-13] &  -3.0 & -3.3 \\
10.0 & 7.86[-14] & 7.12[-14] & 7.10[-14] &  -9.4 & -9.7 \\
\hline
\label{table:trzcomp}
\end{tabular}
\renewcommand{\arraystretch}{1.0}
\end{table}

\begin{table}
\renewcommand{\arraystretch}{0.8}
\caption{Comparison of 10-like total DR rate coefficients from
this work with those of Behar \etal \cite{behar99a}, and the \% difference between the 
two\footnote{The \% difference is calculated as 
$(\alpha_{\mathrm{DR Present}}-\alpha_{\mathrm{DR Behar}})/\alpha_{\mathrm{DR Behar}}$.}.
Quantities in square brackets are powers of ten.}
\begin{tabular}{@{}cccc}
\hline
$T$ (K) &  This work & Behar \etal & \%Diff. \\
\hline
5.80[+05] & 4.19[-22] & 1.71[-21] & -75.5 \\
1.16[+06] & 2.24[-17] & 5.03[-17] & -55.5 \\
2.32[+06] & 1.11[-14] & 1.17[-14] &  -5.4 \\
5.80[+06] & 9.53[-13] & 1.34[-12] & -28.9 \\
1.16[+07] & 4.73[-12] & 6.39[-12] & -26.0 \\
2.32[+07] & 9.66[-12] & 1.12[-11] & -13.8 \\
5.80[+07] & 8.71[-12] & 1.02[-11] & -14.6 \\
1.16[+08] & 5.21[-12] & 6.06[-12] & -14.1 \\
2.32[+08] & 2.55[-12] & 2.83[-12] & -10.0 \\
5.80[+08] & 7.70[-13] & 8.53[-13] &  -9.8 \\
\hline
\label{table:beharcomp}
\end{tabular}
\renewcommand{\arraystretch}{1.0}
\end{table}

\begin{table}
\renewcommand{\arraystretch}{0.8}
\caption{Comparison of 10-like total DR rate coefficients from
 this work with those of Safronova \etal \cite{usafronova09b}. Quantities in square brackets are powers of ten.}
\begin{tabular}{@{}cccc}
\hline
$T$ (K) & This work &  Safronova \etal & \%Diff. \\
\hline
6.30[+05] & 2.54[-21] & 7.67[-21] & -66.9 \\
8.19[+05] & 7.80[-19] & 6.66[-19] &  17.1 \\
1.06[+06] & 9.71[-18] & 1.97[-17] & -50.7 \\
1.38[+06] & 1.20[-16] & 2.61[-16] & -54.0 \\
1.80[+06] & 1.54[-15] & 2.01[-15] & -23.6 \\
2.34[+06] & 1.17[-14] & 1.14[-14] &   2.8 \\
3.04[+06] & 5.23[-14] & 5.40[-14] &  -3.2 \\
3.96[+06] & 2.38[-13] & 2.12[-13] &  12.3 \\
5.14[+06] & 6.31[-13] & 6.47[-13] &  -2.5 \\
6.68[+06] & 1.54[-12] & 1.53[-12] &   0.8 \\
8.68[+06] & 3.31[-12] & 2.86[-12] &  15.7 \\
1.13[+07] & 4.57[-12] & 4.42[-12] &   3.4 \\
1.46[+07] & 6.28[-12] & 5.89[-12] &   6.6 \\
1.90[+07] & 8.68[-12] & 7.00[-12] &  24.0 \\
2.48[+07] & 9.74[-12] & 7.60[-12] &  28.1 \\
3.23[+07] & 1.01[-11] & 7.68[-12] &  31.0 \\
4.19[+07] & 1.03[-11] & 7.29[-12] &  40.6 \\
5.45[+07] & 8.99[-12] & 6.54[-12] &  37.4 \\
7.08[+07] & 7.89[-12] & 5.58[-12] &  41.4 \\
9.21[+07] & 6.53[-12] & 4.56[-12] &  43.2 \\
1.20[+08] & 5.06[-12] & 3.59[-12] &  40.9 \\
1.56[+08] & 3.91[-12] & 2.73[-12] &  43.1 \\
2.02[+08] & 3.02[-12] & 2.03[-12] &  48.8 \\
2.63[+08] & 2.17[-12] & 1.48[-12] &  46.6 \\
3.42[+08] & 1.56[-12] & 1.06[-12] &  47.1 \\
4.44[+08] & 1.11[-12] & 7.47[-13] &  48.8 \\
5.78[+08] & 7.74[-13] & 5.22[-13] &  48.3 \\
7.51[+08] & 5.39[-13] & 3.62[-13] &  49.0 \\
9.76[+08] & 3.72[-13] & 2.50[-13] &  48.7 \\
1.26[+09] & 2.56[-13] & 1.71[-13] &  49.6 \\
1.65[+09] & 1.75[-13] & 1.17[-13] &  49.4 \\
2.15[+09] & 1.19[-13] & 7.96[-14] &  49.8 \\
2.79[+09] & 8.13[-14] & 5.41[-14] &  50.3 \\
3.62[+09] & 5.52[-14] & 3.67[-14] &  50.5 \\
4.71[+09] & 3.74[-14] & 2.49[-14] &  50.0 \\
6.13[+09] & 2.52[-14] & 1.69[-14] &  49.2 \\
\hline
\label{table:safronovacomp}
\end{tabular}
\renewcommand{\arraystretch}{1.0}
\end{table}

\begin{table}
\renewcommand{\arraystretch}{0.8}
\caption{Comparison of 18-like total DR rate coefficients from
 this work with those of Peleg \etal\cite{peleg98a}. Quantities in square brackets are powers of ten.}
\begin{tabular}{@{}cccc}
\hline
$T$ (K) & This work & Peleg \etal & \%Diff. \\
\hline
1.16[+05] & 2.81[-09] & 4.50[-09] & -37.5 \\
2.32[+05] & 2.08[-09] & 3.32[-09] & -37.2 \\
3.48[+05] & 1.81[-09] & 2.68[-09] & -32.5 \\
5.80[+05] & 1.52[-09] & 2.06[-09] & -26.4 \\
1.16[+06] & 1.16[-09] & 1.45[-09] & -19.7 \\
2.32[+06] & 8.23[-10] & 9.51[-10] & -13.4 \\
3.48[+06] & 6.35[-10] & 7.03[-10] &  -9.7 \\
5.80[+06] & 4.24[-10] & 4.59[-10] &  -7.6 \\
1.16[+07] & 2.24[-10] & 2.45[-10] &  -8.4 \\
2.32[+07] & 1.11[-10] & 1.21[-10] &  -8.1 \\
3.48[+07] & 7.10[-11] & 7.72[-11] &  -8.1 \\
5.80[+07] & 3.84[-11] & 4.18[-11] &  -8.1 \\
8.12[+07] & 2.48[-11] & 2.72[-11] &  -8.7 \\
1.16[+08] & 1.55[-11] & 1.70[-11] &  -8.9 \\
2.32[+08] & 5.98[-12] & 6.52[-12] &  -8.3 \\
3.48[+08] & 3.36[-12] & 3.65[-12] &  -7.9 \\
5.80[+08] & 1.60[-12] & 1.74[-12] &  -8.3 \\
\hline
\label{table:pelegcomp}
\end{tabular}
\renewcommand{\arraystretch}{1.0}
\end{table}

\end{document}